\documentclass[a4paper]{article}

\usepackage[latin1]{inputenc}
\usepackage{amsfonts}
\usepackage{amsmath}
\usepackage{amssymb}
\usepackage[UKenglish]{babel}
\usepackage{fontenc}
\usepackage{graphicx}
\usepackage{xypic}
\usepackage{amscd}
\usepackage{psfrag}
\usepackage{amscd}

\newcommand{\CC}{\ensuremath{\mathbb{C}}}
\newcommand{\RR}{\ensuremath{\mathbb{R}}}
\newcommand{\NN}{\ensuremath{\mathbb{N}}}
\newcommand{\ZZ}{\ensuremath{\mathbb{Z}}}

\newcommand{\CP}[1]{\ensuremath{\mathbb{CP}^{#1}}}

\renewcommand{\d}{\ensuremath{\mathrm{d}}}
\newcommand{\covd}{\ensuremath{\nabla}}
\newcommand{\mom}{\ensuremath{\mathfrak m}}

\newcommand{\Ob}[1]{\ensuremath{\mathcal{O}(#1)}} 		
\newcommand{\epsi}{\varepsilon}

\newcommand{\GLi}[1]{\ensuremath{\mathrm{GL}(#1)}}
\newcommand{\Aut}{\ensuremath{\mathrm{Aut}}}

\newcommand{\SU}[1]{\ensuremath{\mathrm{SU}(#1)}}

\newcommand{\PSU}[1]{\ensuremath{\mathrm{PSU}(#1)}}

\newcommand{\U}[1]{\ensuremath{\mathrm{U}(#1)}}
\renewcommand{\u}[1]{\ensuremath{\mathfrak{u}(#1)}}

\newcommand{\Mat}[3]{\ensuremath{\mathrm{M}_{#1\times #2}(#3)}}

\newcommand{\im}{\ensuremath{\bold{i}}}
\newcommand{\half}{\ensuremath{\frac{1}{2}}}
\newcommand{\quart}{\ensuremath{\frac{1}{4}}}
\newcommand{\ihalf}{\ensuremath{\frac{\im}{2}}}

\newcommand{\w}{\wedge}

\renewcommand{\Re}[1]{\ensuremath{\mathrm{Re}}\!\left(#1\right)}
\renewcommand{\Im}[1]{\ensuremath{\mathrm{Im}}\!\left(#1\right)}
\newcommand{\e}{\ensuremath{\mathrm{e}}}

\renewcommand{\dim}[1]{\ensuremath{\mathrm{dim}(#1)}}

\renewcommand{\det}[1]{\ensuremath{\mathrm{det}(#1)}}
\newcommand{\tr}[1]{\ensuremath{\mathrm{tr} \! \left( #1 \right)} }
\newcommand{\Ad}[1]{\ensuremath{\mathrm{Ad}(#1)}}

\newcommand{\Grass}[2]{\ensuremath{\mathrm{Gr}_{#1}(#2)}}

\newcommand{\dvol}{\ensuremath{\mathrm{dvol}}}

\newcommand{\al}{\alpha}

\newcommand{\de}{\delta}
\newcommand{\ze}{\zeta}

\newcommand{\vol}[1]{\ensuremath{\mathrm{Vol}(#1)}}

\newcommand{\gaugeLA}{\ensuremath{\mathfrak{g}}}

\newcommand{\curlE}{\ensuremath{\mathcal{E}}}
\newcommand{\curlH}{\ensuremath{\mathcal{H}}}

\newcommand{\curlN}{\ensuremath{\mathcal{N}}}
\newcommand{\curlM}{\ensuremath{\mathcal{M}}}
\newcommand{\curlT}{\ensuremath{\mathcal{T}}}

\newcommand{\curlP}{\ensuremath{\mathcal{P}}}
\newcommand{\curlZ}{\ensuremath{\mathcal{Z}}}

\newcommand{\Hc}{\ensuremath{\mathrm{H}}}

\newcommand{\pd}{\ensuremath{\partial}}

\newcommand{\dgr}{\dagger}

\newcommand{\ph}{\phantom}

\begin{document}

\pagestyle{plain}

\title{
\vskip -70pt
\begin{flushright}
{\normalsize DAMTP-2012-69} \\
\end{flushright}
\vskip 50pt
{\bf \Large Non-abelian vortices on $\CP{1}$ and Grassmannians}
\vskip 30pt
}

\author{
Norman A. Rink \\
{\small{N.A.Rink@damtp.cam.ac.uk}} \\ \\ 
{\sl Department of Applied Mathematics and Theoretical Physics,}\\
{\sl University of Cambridge,}\\
{\sl Wilberforce Road, Cambridge CB3 0WA, England.}\\
}

\vskip 20pt
\date{8 April 2013}
\vskip 20pt

\maketitle

\begin{abstract}
Many properties of the moduli space of abelian vortices on a compact Riemann surface $\Sigma$ are known. For non-abelian vortices the moduli space is less well understood. Here we consider non-abelian vortices on the Riemann sphere $\CP{1}$, and we study their moduli spaces near the Bradlow limit. We give an explicit description of the moduli space as a K\"ahler quotient of a finite-dimensional linear space. The dimensions of some of these moduli spaces are derived. Strikingly, there exist non-abelian vortex configurations on $\CP{1}$, with non-trivial vortex number, for which the moduli space is a point. This is in stark contrast to the moduli space of abelian vortices.

For a special class of non-abelian vortices the moduli space is a Grassmannian, and the metric near the Bradlow limit is a natural generalization of the Fubini--Study metric on complex projective space. We use this metric to investigate the statistical mechanics of non-abelian vortices. The partition function is found to be analogous to the one for abelian vortices.
\end{abstract}

\newpage
\tableofcontents
\newpage

\section{Introduction}

Gauged vortices with abelian gauge group are a time-honoured subject (see e.g.~\cite{Witten, Nielsen:Olesen, Bradlow:line:bundles, GP:direct, Manton:Sutcliffe, Popov:int, Krusch:Speight, Manton:Rink:hyperbolic}), and various models with non-abelian gauge groups have been studied (e.g.~\cite{GP:1994, Hanany:Tong, Auzzi:et:al, Popov:quiver, Popov:nonab, Baptista:nonabelian, Manton:Sakai, Manton:Rink:nonab}). The term non-abelian vortex refers to a solution of the first order Bogomolny equations in a model with non-abelian gauge group. Quite naturally the intuition we have for the behaviour of non-abelian vortices is much less developed than in the abelian case: The only degrees of freedom of abelian vortices in two dimensions are the coordinates of their centres; they have no internal degrees of freedom. Based on this, one would expect that the degrees of freedom associated with a non-abelian vortex are comprised of its spatial coordinates, as in the abelian case, and additional internal degrees of freedom. The latter are expected to capture the fields' orientations within suitable representations of the gauge algebra. In this paper we study a model on $\CP{1}$ which accommodates non-abelian vortices for which this separation into spatial and internal degrees of freedom breaks down. In such a situation vortices cannot be localized, and therefore it is perhaps unjustified to speak of individual vortices. Nonetheless, we continue to refer to solutions of the relevant Bogomolny equations as vortices, by analogy with the abelian Higgs model.

The model we consider was derived in \cite{Manton:Rink:nonab}, where it was obtained from pure Yang--Mills theory by a symmetry reduction: Starting with Yang--Mills theory on the euclidean background $\Sigma\times S^2$, with $\Sigma$ a Riemann surface, and imposing invariance under rotations of $S^2$ reduces the Yang--Mills theory to a Yang--Mills--Higgs model on $\Sigma$. The self-duality equation of Yang--Mills theory reduces to Bogomolny equations in the Yang--Mills--Higgs model on $\Sigma$. Similar symmetry reductions of Yang--Mills theory have been looked at in many places both in the mathematics \cite{GP:1994, Bradlow:GP:stable, Bradlow:GP:Gothen} and the physics literature \cite{Witten, Popov:quiver, Popov:nonab, Manton:Sakai, Strachan:instantons}. The novelty of the derivation in \cite{Manton:Rink:nonab} lies in the choice of Yang--Mills gauge group, which was taken to be $\PSU{N}$, i.e.~$\SU{N}$ modulo its centre. 

The Bogomolny equations in \cite{Manton:Rink:nonab} are expected to be integrable, provided $\Sigma$ is equipped with a hyperbolic metric, i.e.~has constant negative curvature. This is a consequence of the integrability of the self-dual Yang--Mills equations, see \cite{Popov:int, Popov:nonab}. For the Bogomolny equations of the abelian Higgs model some explicit solutions have been found in special cases \cite{Witten, Krusch:Speight, Manton:Rink:hyperbolic, Dunajski:vortices}, and also in a model with gauge group $\U{1}\times\dots\times\U{1}$ \cite{Strachan:instantons}. However, there is a fundamental mathematical problem that impedes finding explicit solutions if $\Sigma$ has non-trivial topology: Part of such a solution is an explicit expression for the Higgs field, which can be regarded as a holomorphic section of a vector bundle over $\Sigma$. The sections of holomorphic line bundles over an arbitrary Riemann surface $\Sigma$ are not generally known in closed form. Even the number of linearly independent sections may be unknown. Needless to say that the situation is even more complicated for vector bundles.

If, however, one is prepared to give up integrability, one can consider the Bogomolny equations on the Riemann sphere $\CP{1}$. The advantage of this is that the Grothendieck Lemma (see e.g.~chapter 5 in \cite{Huybrechts}) serves to classify holomorphic vector bundles over $\CP{1}$. Moreover, holomorphic sections can be regarded as vectors whose entries are homogeneous polynomials in the homogeneous coordinates of $\CP{1}$. This gives an explicit description of the Higgs field, and the coefficients of the polynomials are the moduli. For abelian vortices this idea was used in \cite{Baptista:Manton:S2} to derive the moduli space metric near the Bradlow limit \cite{Bradlow:line:bundles}. In this paper we generalize the work in \cite{Baptista:Manton:S2} to non-abelian vortices. Near the Bradlow limit the Bogomolny equations reduce to algebraic constraints. Together with a gauge fixing procedure these algebraic constraints give a description of the vortex moduli space as a K\"ahler quotient (cf.~\cite{GP:direct, GP:1994, Hitchin:et:al} and appendix B of \cite{Samols}). The K\"ahler quotient description also equips the moduli space with a natural metric, and this metric agrees with the one whose geodesics describe the slow motion of vortices \cite{Manton:BPS:1982}. In an interesting special case the moduli space is a Grassmannian, and the moduli space metric is a natural generalization of the Fubini--Study metric on complex projective space. This allows us to calculate the volume of the moduli space and hence to study its statistical mechanics, generalizing \cite{Manton:stat:mech}. The volume of the moduli space of non-abelian vortices in a different model was recently calculated in \cite{Miyake:Ohta:Sakai}.

This work is structured as follows: In section \ref{sec:model} we review the model that was introduced in \cite{Manton:Rink:nonab} and set up notation. In sections \ref{sec:near:Bradlow} and \ref{sec:symplectic:quotient} we identify the moduli space of non-abelian vortices near the Bradlow limit as a K\"ahler quotient, and we give a semi-explicit expression for its metric. We derive the dimensions of possible moduli spaces in section \ref{sec:dimensions:examples}. In a special case the moduli space is shown to be a Grassmannian. In this case the moduli space metric can be given explicitly, which we do in section \ref{sec:Grass:applications}. The volume of the moduli space is also derived, and this is used to study the statistical mechanics of non-abelian vortices. Section \ref{sec:summary:outlook} summarizes our results and suggests directions for future work.

\section{A non-abelian Yang--Mills--Higgs model} \label{sec:model}

In this section we review the Yang--Mills--Higgs model from \cite{Manton:Rink:nonab}, with particular emphasis on the geometric structures involved. Our notation is largely the same as in \cite{Manton:Rink:nonab}. 

Let $P$ be a principal bundle on the Riemann surface $\Sigma$ with structure group $G = {\rm S}(\U{m}\!\times\!\U{n})/\ze_N$, where $N = m+n$. The leading ${\rm S}$ means that the overall determinant is one, and $\ze_N$ denotes the centre of ${\rm S}(\U{m}\!\times\!\U{n})$, which is the cyclic group of order $N$. By $\gaugeLA$ we denote the Lie algebra of $G$, i.e.~$\gaugeLA={\rm s}(\u{m}\!\times\!\u{n})$. A unitary connection on $P$ is locally given by the gauge potential $A$,
\begin{align}
 A = \left(\begin{array}{cc} a & 0 \\ 0 & b \end{array}\right). \label{eq:connection:decomp}
\end{align}
Here $a$ and $b$ are locally defined 1-forms with values in the Lie algebras $\u{m}$ and $\u{n}$ respectively, and such that $\tr{a}+\tr{b}=0$. The corresponding curvature $F^A$ decomposes accordingly,
\begin{align}
 F^A = \d A + A\w A = \left(\begin{array}{cc} \d a + a\w a & 0 \\ 0 & \d b + b\w b \end{array}\right) = \left(\begin{array}{cc} f^a & 0 \\ 0 & f^b \end{array}\right), \label{eq:curvature:decomp}
\end{align}
where the last equality defines the curvatures $f^a$, $f^b$.

The gauge group $G$ acts on $\Mat{N}{N}{\CC}$, the space of $N\!\times\!N$ matrices with complex entries, by the adjoint representation of $\U{N}$,
\begin{align}
 M \mapsto \Ad{g}M = g M g^{-1}, \quad M\in\Mat{N}{N}{\CC}, \quad g\in\U{N}.
\end{align}
This action of $\U{N}$ does not depend on elements in the centre $\ze_N$, and hence yields an action of $G$. Therefore we can introduce on $\Sigma$ the vector bundle associated to $P$ via the representation ${\rm Ad}$,
\begin{align}
 E = P \times_{\rm Ad} \Mat{N}{N}{\CC}. \label{eq:bundle:E:def}
\end{align}
The bundle $E$ inherits a covariant derivative from the connection $A$,
\begin{align}
 D^E = \d + [A,\cdot\,].
\end{align}
For $g\in G$ choose the representative
\begin{align}
 g = \left(\begin{array}{cc} g_a & 0 \\ 0 & g_b \end{array}\right)\,\e^{2\pi\im\frac{k}{N}}, \quad g_a\in\U{m}, \quad g_b\in\U{n}, \quad k\in\NN, \label{eq:gauge:rep}
\end{align}
and write $M\in\Mat{N}{N}{\CC}$ as
\begin{align}
 M = \left(\begin{array}{cc} M_{11} & M_{12} \\ M_{21} & M_{22} \end{array}\right),
\end{align}
with $M_{11}\in\Mat{m}{m}{\CC}$, $M_{12}\in\Mat{m}{n}{\CC}$, $M_{21}\in\Mat{n}{m}{\CC}$, $M_{22}\in\Mat{n}{n}{\CC}$. It follows from
\begin{align}
 \Ad{g}M = \left(\begin{array}{cc} g_aM_{11}g_a^{-1} & g_aM_{12}g_b^{-1} \\ g_bM_{21}g_a^{-1} & g_bM_{22}g_b^{-1} \end{array}\right), 
\end{align}
that the bundle $E$ decomposes as
\begin{align}
 E = E_{11} \oplus E_{12} \oplus E_{21} \oplus E_{22}. \label{eq:E:decomposition}
\end{align}
The symmetry reduction in \cite{Manton:Rink:nonab} leads to a Higgs field $\phi$ which is a section of $E_{21}$. Restricting the covariant derivative $D^E$ to $E_{21}$ yields
\begin{align}
 D\phi = \d\phi + b\phi - \phi a. \label{eq:covar:deriv}
\end{align}
Consequently $\phi^{\dgr}$ is a section of $E_{12}$, with covariant derivative
\begin{align}
 D\phi^{\dgr} = \d\phi^{\dgr} + a\phi^{\dgr} - \phi^{\dgr}b.
\end{align}

To write down an energy functional for the Yang--Mills--Higgs model and the corresponding Bogomolny equations, we need a metric on $\Sigma$. We take
\begin{align}
 \d s^2 = \Omega(x^1,x^2) ((\d{x^1})^2 + (\d{x^1})^2) = \Omega(z,\bar{z}) \d{z}\d{\bar z}, 
\end{align}
where $z$ is a local complex coordinate on $\Sigma$, and the real coordinates $x^1$, $x^2$ are defined by $z=x^1+\im x^2$, ${\bar z}=x^1-\im x^2$. Since $\Sigma$ has real dimension two, the K\"ahler form $\omega_{\Sigma}$ and the volume form agree,
\begin{align}
 \omega_{\Sigma} = \ihalf \Omega(z,\bar{z}) \d{z}\w\d{\bar z} = \dvol_{\Sigma} ,
\end{align}
For future reference we also give explicitly the Hodge $*$ operator on functions, 1-forms, and 2-forms,
\begin{align}
 &*f = f\omega_{\Sigma}, \\
 &*(\al_z\d{z} + \al_{\bar z}\d{\bar z}) = -\im\al_z\d{z} + \im\al_{\bar z}\d{\bar z}, \\
 &*(\eta_{z\bar z} \d{z}\w\d{\bar z}) = -2\im\Omega^{-1}\eta_{z\bar z} = \frac{\eta_{z\bar z} \d{z}\w\d{\bar z}}{\omega_{\Sigma}}. 
\end{align}
Note in particular that on 1-forms $*^2 = -1$. From this it follows that $*$ is a complex structure on the space of connections on $P$. We will use this observation at the end of subsection \ref{sec:moduli:space:metric}.

The energy functional of the Yang--Mills--Higgs model derived in \cite{Manton:Rink:nonab} is
\begin{align}
 E_{\rm 2d} = \half \int_{\Sigma} &\left( -\tr{f^a\w*f^a} -\tr{f^b\w*f^b}  
                         +\half\tr{D\phi\w*D\phi^{\dgr}} \ph{\frac{1}{8}} \right. \nonumber \\ 
                         &\ph{=}\left. +\frac{1}{8}\tr{\mathbb{I}_n-\phi\phi^{\dgr}}^2\omega_{\Sigma} \right) 
			 + \frac{1}{16}\frac{n(m-n)}{N}\vol{\Sigma}, \label{eq:energy}
\end{align}
where the minus signs occur in front of the Yang--Mills kinetic terms since $f^a$ and $f^b$ are anti-hermitian. The constant term proportional to $\vol{\Sigma}$ was shown in \cite{Manton:Rink:nonab} to be a consequence of the fact that $E_{\rm 2d}$ is obtained from the Yang--Mills action in four dimensions by a symmetry reduction. A Bogomolny argument can be carried out on the energy functional \eqref{eq:energy}, leading to the Bogomolny equations
\begin{align}
 &D_{\bar z}\phi = 0 \label{eq:Bog:1}, \\
 &f^a_{z \bar z} = \frac{\Omega}{8} \left( -\frac{2n}{N}\mathbb{I}_m + \phi^{\dgr} \phi\right), \label{eq:Bog:2}\\
 &f^b_{z \bar z} = \frac{\Omega}{8} \left(  \frac{2m}{N}\mathbb{I}_n - \phi \phi^{\dgr}\right). \label{eq:Bog:3}
\end{align}
It was shown in \cite{Manton:Rink:nonab} that when the Bogomolny equations are satisfied, the energy functional \eqref{eq:energy} reduces to
\begin{align}
 \curlE = \frac{\pi}{N}c_1(E_{21}), \label{eq:energy:1st:Chern}
\end{align}
where $c_1$ denotes the first Chern number. We refer to solutions of \eqref{eq:Bog:1}-\eqref{eq:Bog:3} as non-abelian vortices, and we call $c_1(E_{21})$ the non-abelian vortex number. This is justified since for $m=n=1$ the equations \eqref{eq:Bog:1}-\eqref{eq:Bog:3} reduce to the Bogomolny equations of the abelian Higgs model \cite{note:traceless}.

Solutions of \eqref{eq:Bog:1}-\eqref{eq:Bog:3} are physically equivalent if they are related by a gauge transformation. We are interested in the moduli space of non-abelian vortices, which is the space of solutions of \eqref{eq:Bog:1}-\eqref{eq:Bog:3} modulo gauge transformations. The moduli space can be described as a K\"ahler quotient \cite{GP:direct, GP:1994} by identifying equations \eqref{eq:Bog:2} and \eqref{eq:Bog:3} as the level set equations of the moment map for the action of gauge transformations. In the next section we show how equations \eqref{eq:Bog:2} and \eqref{eq:Bog:3} reduce to algebraic constraints if $\Sigma=\CP{1}$ and near the Bradlow limit. In section \ref{sec:symplectic:quotient} we identify these constraints with moment maps on finite dimensional linear spaces.

\section{Near the Bradlow limit} \label{sec:near:Bradlow}

From now on we take $\Sigma=\CP{1}$. This allows us to solve explicitly the first Bogomolny equation \eqref{eq:Bog:1}. Namely, the first Bogomolny equation says that $\phi$ is a holomorphic section of $E_{21}$. By the Grothendieck Lemma the vector bundle $E_{21}$ over $\CP{1}$ decomposes into a sum of line bundles,
\begin{align}
 E_{21} = \bigoplus_{i=1}^{mn} \Ob{k_i}, \label{eq:E:Grothendieck}
\end{align}
where $\Ob{k_i}$ denotes the holomorphic line bundle over $\CP{1}$ of degree $k_i\in\ZZ$. It follows that the entries of $\phi$ are sections of the $\Ob{k_i}$, which in turn can be described as homogeneous polynomials of degree $k_i$. 

The Bogomolny equations \eqref{eq:Bog:1}, \eqref{eq:Bog:2} have solutions only if 
\begin{align}
 \vol{\CP{1}} \ge 4\pi \frac{c_1(E_{21})}{mn}. \label{eq:Bradlow:bound}
\end{align}
This is the generalized Bradlow bound for non-abelian vortices, cf.~\cite{Bradlow:line:bundles, Manton:Rink:nonab}. In the strict Bradlow limit, i.e.~when equality holds in \eqref{eq:Bradlow:bound}, the Higgs field vanishes identically, $\phi=0$, and the moduli space consists of a single point, as we shall see. Near the Bradlow limit, when $\vol{\CP{1}}$ slightly exceeds the lower bound \eqref{eq:Bradlow:bound}, the magnitude of the Higgs field $\phi$ is small, and as a consequence the Bogomolny equations simplify. We will take advantage of this to study properties of the moduli space. Since increasing $\vol{\CP{1}}$ is a smooth process, statements about the topological properties of the moduli space near the Bradlow limit are expected to hold for general values of $\vol{\CP{1}}$.

We now briefly introduce our conventions regarding $\CP{1}$. Thinking of $\CP{1}$ as the sphere $S^2$, it can be covered with two open sets, $U_0$ and $U_1$, where $U_0$ consist of all points of $S^2$ except the north pole, and $U_1$ consists of all points except the south pole. We denote the complex coordinate on $U_0$ as $z$, and the one on $U_1$ as $z'$. The coordinate $z$ is obtained by stereographic projection from the north pole onto the equatorial plane, and $z'$ is obtained analogously by projecting from the south pole. On $U_0\cap U_1$ the local coordinates satisfy $z'=1/z$. For $k\in\ZZ$ the holomorphic transition function
\begin{align}
 &g_{01} \colon U_0\cap U_1 \to \CC^*, \\
 &g_{01}(z) = z^{k},	
\end{align}
defines the holomorphic line bundle $\Ob{k}$ of degree $k$ over $\CP{1}$. Unless otherwise stated, we always work on the open set $U_0$.  We equip $\CP{1}$ with the standard round metric given by the conformal factor 
\begin{align}
 \Omega(z,\bar z) = \frac{4R^2}{(1+z\bar z)^2},
\end{align}
where $R$ is the radius of $\CP{1}$ when regarded as the sphere $S^2$. The corresponding K\"ahler form,
\begin{align}
 \omega_{\CP{1}}  = \ihalf \frac{4R^2}{(1+z\bar z)^2} \d{z}\w\d{\bar z} ,
\end{align}
is a multiple of the Fubini--Study form on $\CP{1}$. The area of $\CP{1}$ is
\begin{align}
 \vol{\CP{1}} = \int_{\CP{1}} \omega_{\CP{1}} = 4\pi R^2.
\end{align}

To solve the Bogomolny equations, it is convenient to work in holomorphic gauge: The gauge potential $A$ can be expanded in its 1-form components,
\begin{align}
 A = A_z \d{z} + A_{\bar z}\d{\bar z}.
\end{align}
Holomorphic gauge is defined by the condition $A_{\bar z} = 0$. In \cite{Manton:Rink:hyperbolic} it was explained how to go to holomorphic gauge in the abelian Higgs model. The procedure is completely analogous in the non-abelian model we are studying here: First one introduces a hermitian structure $h$ on the bundle $P$. In unitary gauge one has $h = \mathbb{I}_N$, and the gauge group of $P$ is ${\rm S}(\U{m}\times\U{n})/\ze_N$. In a general gauge $h$ is locally given by positive definite hermitian matrices of the form
\begin{align}
 h = \left(\begin{array}{cc} h^a & 0 \\ 0 & h^b \end{array}\right), \quad h^a\in\Mat{m}{m}{\CC}, \: h^b\in\Mat{n}{n}{\CC}, \label{eq:hermitian:structure}
\end{align}
and the structure group of $P$ is ${\rm S}(\GLi{m}\times\GLi{n})/\ze_N$. In unitary gauge the connection $A$ on $P$ satisfies
\begin{align}
 A^{\dgr} = -A.
\end{align}
In a general gauge this equation is replaced by the compatibility condition
\begin{align}
 \d h = A^{\dgr} h + h A.
\end{align}
Therefore, in holomorphic gauge the condition $A_{\bar z} = 0$ fully determines $A$ in terms of the hermitian structure $h$,
\begin{align}
 A_z = h^{-1} \pd_z h,
\end{align}
and this is known as the Chern connection, see e.g.~\cite{Huybrechts, Moroianu}. It follows that
\begin{align}
 &a_z = (h^{a})^{-1}\pd_z h^a, \quad a_{\bar z} = 0, \\
 &b_z = (h^{b})^{-1}\pd_z h^b, \quad b_{\bar z} = 0. 
\end{align}
On $\CP{1}$ and in holomorphic gauge the Bogomolny equations \eqref{eq:Bog:1}-\eqref{eq:Bog:3} read
\begin{align}
 &\pd_{\bar z}\phi = 0, \label{eq:holom:Bog:1} \\
 &f^a_{z\bar z} = \frac{R^2}{2(1+z\bar z)^2} \left( - \frac{2n}{N}\mathbb{I}_m + (h^a)^{-1}\phi^{\dgr}h^b\phi\right), \label{eq:holom:Bog:2} \\
 &f^b_{z\bar z} = \frac{R^2}{2(1+z\bar z)^2} \left(   \frac{2m}{N}\mathbb{I}_n - \phi (h^a)^{-1}\phi^{\dgr}h^b\right), \label{eq:holom:Bog:3} 
\end{align}
with the field strengths
\begin{align}
 &f^a_{z\bar z} = -\pd_{\bar z}((h^a)^{-1}\pd_z h^a), \label{eq:metric:curvature:a} \\
 &f^b_{z\bar z} = -\pd_{\bar z}((h^b)^{-1}\pd_z h^b). \label{eq:metric:curvature:b}
\end{align}

\subsection{The dissolved limit} \label{sec:dissolved}

Using the terminology of \cite{Manton:Romao:Jac}, we also refer to the Bradlow limit as the dissolved limit. This is because abelian vortices are centred at the zeros of $\phi$, and vortices are fully dissolved in the Bradlow limit since $\phi=0$ identically. 

We make the following ansatz for the hermitian structure $h$, 
\begin{align}
 &h^a(z,\bar z) = (1+z\bar z)^{d_a} \mathbb{I}_m, \\
 &h^b(z,\bar z) = (1+z\bar z)^{d_b} \mathbb{I}_n.
\end{align}
The exponents $d_a$ and $d_b$ are real constants, and their values must be consistent with the topology of $P$. By this is meant that $h^a$ and $h^b$ must transform under gauge transformations as
\begin{align}
 &h^a \mapsto (g_a^{-1})^{\dgr} h^a g_a^{-1}, \\
 &h^b \mapsto (g_b^{-1})^{\dgr} h^b g_b^{-1}, 
\end{align}
where $g_a$ and $g_b$ are as in \eqref{eq:gauge:rep}. (Note that no index contraction is implied. The symbols $a$ and $b$ merely label the block entries of $g$ in \eqref{eq:gauge:rep} and of $h$ in \eqref{eq:hermitian:structure}.)

Setting $\phi=0$ and using the above ans\"atze for $h^a$, $h^b$, equations \eqref{eq:holom:Bog:2}, \eqref{eq:holom:Bog:3} lead to
\begin{align}
 &d_a = \frac{n}{N} R^2, \\
 &d_b = -\frac{m}{N} R^2.
\end{align}
This does not appear to constrain $d_a$ and $d_b$ in any way since $R$ can take any value. However, in the Bradlow limit the constants $d_a$ and $d_b$ are fixed by virtue of \eqref{eq:Bradlow:bound}. For the vortex number in the Bradlow limit we find
\begin{align}
 c_1(E_{21}) &= \frac{\im}{2\pi}\int_{\CP{1}} \left( \tr{m f^b} -  \tr{n f^a} \right) \\
	     &= mn(d_a-d_b) \\
	     &= mn R^2.
\end{align}
Since $c_1(E_{21})$ is integral, this leads to the constraint
\begin{align}
 R^2 \in \frac{1}{mn} \ZZ,
\end{align}
and we also have
\begin{align}
 &d_a = \frac{c_1(E_{21})}{mN}, \label{eq:constraint:da} \\
 &d_b = -\frac{c_1(E_{21})}{nN}. \label{eq:constraint:db}
\end{align}
We give a special name to the value of $R$ for which the bound \eqref{eq:Bradlow:bound} is saturated,
\begin{align}
 R_B^2 = \frac{c_1(E_{21})}{mn}.
\end{align}
Note that $c_1(E_{mn})=0$ is compatible with the Bradlow limit only if $R_B=0$, i.e.~$\CP{1}$ degenerates to a point. We usually assume $R_B\ne 0$, and we will deal with the case $R_B=0$ separately.

The degrees $k_i$ in \eqref{eq:E:Grothendieck} are uniquely determined in the Bradlow limit. This can be seen by inspecting the hermitian structure on $E_{21}$. For arbitrary smooth sections $\psi_1, \psi_2\in\Gamma(\CP{1},E_{21})$,  
\begin{align}
 (\psi_1,\psi_2)_h = \tr{(h^a)^{-1}\psi_1^{\dgr} h^b \psi_2} = (1+z\bar z)^{d_b-d_a} \tr{\psi_1^{\dgr}\psi_2}.
\end{align}
This implies that $k_i=k=d_a-d_b$ for all $i=1,\dots,mn$. Alternatively, one can employ a more abstract argument to show that all the $k_i$ are identical: The Bogomolny equations \eqref{eq:holom:Bog:2}, \eqref{eq:holom:Bog:3} with $\phi=0$ imply that the vector bundle $E_{21}$ is Einstein--Hermitian. Therefore $E_{21}$ decomposes into a direct sum of stable bundles all of which have the same slope, see section V.2 in \cite{Kobayashi:vector:bundles}. We also refer to \cite{Kobayashi:vector:bundles} for the definitions of Einstein--Hermitian vector bundles, stable bundles, and slope. In the case where $E_{21}$ is an Einstein--Hermitian bundle over $\CP{1}$ it follows again from the Grothendieck Lemma that $E_{21}$ decomposes into line bundles of the same degree $k$. Since $k$ is fully determined by the Bogomolny equations with $\phi=0$, it also follows that in the Bradlow limit the moduli space is a point.

Now, $c_1(E_{21}) = mnk$, and thus \eqref{eq:constraint:da} and \eqref{eq:constraint:db} read
\begin{align}
 &d_a = \frac{nk}{N}, \label{eq:da:k} \\
 &d_b = -\frac{mk}{N}, \label{eq:db:k}
\end{align}
which is consistent with $k=d_a-d_b$. Moreover, the fact that $d_a,d_b\in\frac{1}{N}\ZZ$ is consistent with the structure group of $P$, in holomorphic gauge, being ${\rm S}(\GLi{m}\!\times\!\GLi{n})/\ze_N$ (cf.~\cite{Manton:Rink:nonab}). The local form on $U_0$ of the hermitian structure $h$ on $P$ is 
\begin{align}
 h(z, \bar z) = \left(\begin{array}{cc} (1+z\bar z)^{ \frac{nk}{N}}\mathbb{I}_m & 0 \\ 0 & (1+z\bar z)^{-\frac{mk}{N}}\mathbb{I}_n \end{array}\right).
\end{align}
On $U_1$ the hermitian structure is fully determined by $h(z, \bar z)$ and the transition function
\begin{align}
 g_{01}(z) = \left(\begin{array}{cc} z^{-\frac{nk}{N}}\mathbb{I}_m & 0 \\ 0 & z^{\frac{mk}{N}}\mathbb{I}_n \end{array}\right) \e^{2\pi\im\frac{l}{N}}, 
 \quad l\in\ZZ. \label{eq:transition:g01}
\end{align}
Hence,
\begin{align}
 h(z', \bar z') = \left(\begin{array}{cc} (1+z'\bar z')^{ \frac{nk}{N}}\mathbb{I}_m & 0 \\ 0 & (1+z'\bar z')^{-\frac{mk}{N}}\mathbb{I}_n \end{array}\right).
\end{align}
The expression for $g_{01}$ may require clarification: Determining the value $g_{01}(z)$ requires taking the $N$-th root of $z$, which is defined only up to an element of $\ze_N$. Since the structure group of $P$ is ${\rm S}(\GLi{m}\!\times\!\GLi{n})/\ze_N$, the transition function $g_{01}$ is well-defined if the same representative is chosen for the $N$-th root of $z$ in all entries of $g_{01}$. This can be achieved by defining
\begin{align}
 z^{\frac{1}{N}} = \e^{\frac{1}{N}\log z},
\end{align}
and using a fixed branch of the logarithm. 

The upshot of the discussion in this subsection is that in the Bradlow limit $d_a$ and $d_b$ are determined by \eqref{eq:da:k} and \eqref{eq:db:k}, and the radius $R$ is fixed by the vortex number,
\begin{align}
 &c_1(E_{21}) = mnk, \\ 
 &R^2 = R_B^2 = k.
\end{align}

\subsection{Dissolving vortices} \label{sec:dissolving}

To allow for a nontrivial Higgs field, we must move away from the Bradlow limit by increasing the area of $\CP{1}$. Therefore let
\begin{align}
 R = R_B (1 + \epsi),
\end{align} 
with the dimensionless parameter $\epsi$. For small $\epsi$ the Higgs field will deviate only mildly from zero. In the abelian Higgs model this means that vortices are nearly dissolved when $\epsi$ is small, and we refer to this as the regime of dissolving vortices.

We assume that for small $\epsi$ neither the topology nor the holomorphic structure of $E_{21}$ is changed. For the topology this assumption is justified since $E_{21}$, as a smooth bundle over $\CP{1}$, is fully determined by $c_1(E_{21})$, which is an integer and therefore cannot vary smoothly with $\epsi$. The Grothendieck Lemma says that the moduli space of holomorphic structures on $E_{21}$ is a discrete space. Therefore we do not expect the decomposition
\begin{align}
 E_{21} = \bigoplus_{i=1}^{mn} \Ob{k}
\end{align}
to change with $\epsi$.

By the Bogomolny equation \eqref{eq:holom:Bog:1} the Higgs field $\phi$ is a holomorphic section of $E_{21}$. Holomorphic sections of $\Ob{k}$ are homogeneous polynomials of degree $k$ in the homogeneous coordinates of $\CP{1}$. Therefore, in terms of the inhomogeneous coordinate $z$ on $U_0$,
\begin{align}
 \phi(z) = \sqrt{\epsi}\sum_{r=0}^k V_{r} z^r,
\end{align}
where $V_r\in\Mat{n}{m}{\CC}$ for $r=0, \dots,k$. It is clear from the previous discussion that $\phi$ must vanish for $\epsi=0$. That $\phi$ is proportional to $\sqrt{\epsi}$ for dissolving vortices will become apparent shortly. For completeness we give the corresponding expression for $\phi$ on $U_1$,
\begin{align}
	\phi(z') = \sqrt{\epsi}\sum_{r=0}^k V_{k-r} {z'}^r.
\end{align}

\paragraph{}
We now turn to equations \eqref{eq:holom:Bog:2} and \eqref{eq:holom:Bog:3}. Since $\phi$ is essentially fixed by the holomorphic structure of $E_{21}$, the only freedom lies in the hermitian structure $h$ on $P$. If $M^a$ and $M^b$ are smooth functions on $\CP{1}$ with values in the hermitian matrices, then the following modified ansatz for $h$ is consistent with the topology and holomorphic structure,
\begin{align}
 &h^a(z,\bar z) = (1+z\bar z)^{\frac{nk}{N}}\,e^{-2\epsi M^a(z,\bar z)}, \label{eq:herm:structure:a} \\
 &h^b(z,\bar z) = (1+z\bar z)^{-\frac{mk}{N}}\,\,e^{-2\epsi M^b(z,\bar z)}. 
\end{align}
For $\epsi=0$ we clearly recover the situation described in the previous subsection. Using \eqref{eq:metric:curvature:a} and \eqref{eq:metric:curvature:b}, the curvatures derived from these expressions for $h^a$ and $h^b$ are
\begin{align}
 &f^a_{z\bar z} = - \frac{nk}{N(1+z\bar z)^2} \mathbb{I}_m + 2\epsi\,\pd_{z}\pd_{\bar z} M^a, \label{eq:curv:1} \\
 &f^b_{z\bar z} = \frac{mk}{N(1+z\bar z)^2} \mathbb{I}_n + 2\epsi\,\pd_{z}\pd_{\bar z} M^b, \label{eq:curv:2}
\end{align}
up to first order in $\epsi$. Substituting for $f^a$ and $f^b$ in \eqref{eq:holom:Bog:2}, \eqref{eq:holom:Bog:3}, we obtain
\begin{align}
 - \frac{nk}{N(1+z\bar z)^2} \mathbb{I}_m + 2\epsi\,\pd_{z}\pd_{\bar z} M^a 
 &= - \frac{n R^2}{N(1+z\bar z)^2} \mathbb{I}_m + \frac{R^2}{2(1+z\bar z)^{2+k}} \phi^{\dgr}\phi, \label{eq:aux:Bog:1} \\
 \frac{mk}{N(1+z\bar z)^2} \mathbb{I}_n + 2\epsi\,\pd_{z}\pd_{\bar z} M^b 
 &= \frac{m R^2}{N(1+z\bar z)^2} \mathbb{I}_n - \frac{R^2}{2(1+z\bar z)^{2+k}} \phi\phi^{\dgr}. \label{eq:aux:Bog:2}
\end{align}
Next we use the expansion 
\begin{align}
	R^2 = R_B^2 + 2R_B^2\epsi = k + 2k\epsi,
\end{align}
to cancel terms in \eqref{eq:aux:Bog:1}, \eqref{eq:aux:Bog:2}. Thus, up to first order in $\epsi$,
\begin{align}
 2\epsi\,\pd_{z}\pd_{\bar z} M^a 
 &= - \frac{2nk\epsi}{N(1+z\bar z)^2} \mathbb{I}_m + \frac{k}{2(1+z\bar z)^{2+k}} \phi^{\dgr}\phi, \\
 2\epsi\,\pd_{z}\pd_{\bar z} M^b 
 &= \frac{2mk\epsi}{N(1+z\bar z)^2} \mathbb{I}_n - \frac{k}{2(1+z\bar z)^{2+k}} \phi\phi^{\dgr}, 
\end{align}
where we have used that $\phi^{\dgr}\phi \sim \epsi$, $\phi\phi^{\dgr} \sim \epsi$. If $\phi$ were not proportional to $\sqrt{\epsi}$, the right-hand sides of the above equations would not vanish to the right order as $\epsi\to 0$. Finally, cancelling a factor $\epsi$ and expanding $\phi$ in $z$,
\begin{align}
 \pd_{z}\pd_{\bar z} M_a 
 &= - \frac{nk}{N(1+z\bar z)^2} \mathbb{I}_m + \frac{k}{4(1+z\bar z)^{2+k}} \sum_{r,s=0}^{k} V_r^{\dgr} V_s \,{\bar z}^r z^s, 
 \label{eq:dissing:Bog:1} \\
 \pd_{z}\pd_{\bar z} M_b 
 &= \frac{mk}{N(1+z\bar z)^2} \mathbb{I}_n - \frac{k}{4(1+z\bar z)^{2+k}} \sum_{r,s=0}^{k} V_r V_s^{\dgr} z^r {\bar z}^s.
 \label{eq:dissing:Bog:2}
\end{align}

\paragraph{}
The differential operators on the left-hand sides of \eqref{eq:dissing:Bog:1} and \eqref{eq:dissing:Bog:2} are Laplacians, and we can appeal to Hodge theory (see e.g.~section 6 of chapter 0 in \cite{Griffiths:Harris}) to conclude that these equations have solutions if and only if
\begin{align}
 &\int_{\CP{1}}\pd_{z}\pd_{\bar z} M^a\, \d{z}\w\d{\bar z} = 0, \\
 &\int_{\CP{1}}\pd_{z}\pd_{\bar z} M^b\, \d{z}\w\d{\bar z} = 0.
\end{align}
In order to integrate the right-hand sides of \eqref{eq:dissing:Bog:1}, \eqref{eq:dissing:Bog:2}, we need the formula 
\begin{align}
 \int_{\CP{1}} \frac{{\bar z}^r z^s}{(1+z\bar z)^{2+k}}\d{z}\w\d{\bar z} = (-2\pi\im)\de_{r,s} \frac{r!(k-r)!}{(k+1)!}.
\end{align}
This was derived in \cite{Baptista:Manton:S2}. To simplify notation in what follows, we define
\begin{align}
 I_{r,k} = \frac{r!(k-r)!}{(k+1)!}.
\end{align}
Integrating \eqref{eq:dissing:Bog:1} and \eqref{eq:dissing:Bog:2} thus yields
\begin{align}
 &\frac{n}{N} \mathbb{I}_m = \frac{1}{4} \sum_{r=0}^{k} I_{r,k} V_r^{\dgr} V_r, \label{eq:constraint:Bog1} \\
 &\frac{m}{N} \mathbb{I}_n = \frac{1}{4} \sum_{r=0}^{k} I_{r,k} V_r V_r^{\dgr}. \label{eq:constraint:Bog:2} 
\end{align}
Introducing $W_r = \sqrt{I_{r,k}}\, V_r$, we can write more compactly,
\begin{align}
 &\frac{n}{N} \mathbb{I}_m = \frac{1}{4} \sum_{r=0}^{k} W_r^{\dgr} W_r, \label{eq:W:Bog:1} \\
 &\frac{m}{N} \mathbb{I}_n = \frac{1}{4} \sum_{r=0}^{k} W_r W_r^{\dgr}. \label{eq:W:Bog:2} 
\end{align}
These are the Bogomolny equations for dissolving non-abelian vortices on $\CP{1}$. We see that \eqref{eq:W:Bog:1}, \eqref{eq:W:Bog:2} are purely algebraic constraints on the matrices $W_r$, as claimed in the introduction. In section \ref{sec:symplectic:quotient} we will identify these constraints as the defining equations for the level sets of two moment maps on the space $(\Mat{n}{m}{\CC})^{\oplus(k+1)}$.

\subsection{The case $k=0$ and the small sphere}

We already noted in subsection \ref{sec:dissolved} that $c_1(E_{mn})=0$ is consistent with the Bradlow limit only if $\CP{1}$ shrinks to a point. Here we analyze the regime of dissolving vortices with $k=0$ and where the area of $\CP{1}$ is small. To this end, we take the radius of $\CP{1}$ to be $\epsi l$, where $l$ is an arbitrary length scale and $\epsi$ is dimensionless and small, as before. This situation we refer to as the small sphere regime.

In the Bradlow limit the bundle $P$ is necessarily trivial since it is a bundle over a point. We assume that triviality of $P$, and hence of the bundle $E_{21}$, persists in the small sphere regime. Thus the hermitian structures $h^a$ and $h^b$ are functions with values in the positive definite hermitian matrices, 
\begin{align}
 &h^a(z,\bar z) = e^{-2\epsi M^a(z,\bar z)}, \\
 &h^b(z,\bar z) = e^{-2\epsi M^b(z,\bar z)}.
\end{align}
These expressions are consistent with the expressions for $h^a$, $h^b$ in the previous subsection, setting $k=0$. The Bogomolny equations \eqref{eq:holom:Bog:2}, \eqref{eq:holom:Bog:3}, up to second order in $\epsi$, now read
\begin{align}
 2\epsi\,\pd_{z}\pd_{\bar z} M^a 
 &= - \frac{n \epsi^2l^2}{N(1+z\bar z)^2} \mathbb{I}_m + \frac{\epsi^2l^2}{2(1+z\bar z)^2} \phi^{\dgr}\phi, \label{eq:small:Bog:1} \\
 2\epsi\,\pd_{z}\pd_{\bar z} M^b 
 &= \frac{m \epsi^2l^2}{N(1+z\bar z)^2} \mathbb{I}_n - \frac{\epsi^2l^2}{2(1+z\bar z)^2} \phi\phi^{\dgr}. \label{eq:small:Bog:2} 
\end{align}
Since the left-hand and right-hand sides are of different orders in $\epsi$, they must vanish individually. This implies that $M^a$ and $M^b$ are harmonic functions on $\CP{1}$, and hence constants. The Higgs field, being a holomorphic section of the trivial bundle, i.e.~a holomorphic function, is necessarily constant too,
\begin{align}
 \phi = V_0 \in\Mat{n}{m}{\CC}.
\end{align}
Then the right-hand sides of \eqref{eq:small:Bog:1} and \eqref{eq:small:Bog:2} yield
\begin{align}
 &\frac{n}{N} \mathbb{I}_m = \frac{1}{2} V_0^{\dgr} V_0, \label{eq:V0:Bog:1}  \\
 &\frac{m}{N} \mathbb{I}_n = \frac{1}{2} V_0 V_0^{\dgr}. \label{eq:V0:Bog:2}
\end{align}
Note that equations \eqref{eq:V0:Bog:1} and \eqref{eq:V0:Bog:2} are formally nearly identical with \eqref{eq:W:Bog:1} and \eqref{eq:W:Bog:2}.

\subsection{Holomorphic gauge transformations}

We have seen that dissolving vortices are given by a collection of matrices $(W_r)_{r=0,\dots,k}$ that satisfy \eqref{eq:W:Bog:1} and \eqref{eq:W:Bog:2}. To obtain the moduli space, we must identify solutions that are related by gauge transformations. We now determine which gauge transformations are allowed in holomorphic gauge.

A holomorphic gauge transformation $g\in\Aut(P)$ is given by two holomorphic maps $g_0\colon U_0 \to {\rm S}(\GLi{m}\!\times\!\GLi{n})/\ze_N$, $g_1\colon U_1 \to {\rm S}(\GLi{m}\!\times\!\GLi{n})/\ze_N$, such that
\begin{align}
 g_1 = g_{01} g_0 g_{01}^{-1}, \quad\text{on } U_0\cap U_1
\end{align}
where $g_{01}$ is as in \eqref{eq:transition:g01}. Since $g_{01}$ is diagonal and $g_0$ is block-diagonal, we have in fact 
\begin{align}
	g_1 = g_0, \quad\text{on } U_0\cap U_1.
\end{align}
Thus a holomorphic gauge transformation is a globally defined holomorphic function on $\CP{1}$, and hence is constant. We therefore omit the subscripts $0$ and $1$ on maps defining gauge transformations, and we write
\begin{align}
	g = \left(\begin{array}{cc} g_a & 0 \\ 0 & g_b \end{array}\right) \e^{2\pi\im\frac{l}{N}}, \quad l\in\ZZ,
\end{align}
where $\det{g_a}\det{g_b}=1$. It follows that
\begin{align}
 \left\{ g\in \Aut(P) \colon g\text{ is holomorphic} \right\} = {\rm S}(\GLi{m}\!\times\!\GLi{n})/\ze_N.
\end{align}

\paragraph{}
The fields in \eqref{eq:holom:Bog:1}, \eqref{eq:holom:Bog:2} are acted upon by gauge transformations as follows,
\begin{align}
 &h^a \mapsto (g_a^{-1})^{\dgr}h^a\,g_a^{-1}, \label{eq:metric:a:trafo} \\
 &h^b \mapsto (g_b^{-1})^{\dgr}h^b\,g_b^{-1}, \label{eq:metric:b:trafo} \\ 
 &\phi \mapsto g_b\, \phi\, g_a^{-1},
\end{align}
and consequently,
\begin{align}
 &f^a \mapsto g_a\,f^a g_a^{-1},\\
 &f^b \mapsto g_b\,f^b g_b^{-1}.
\end{align}
Expression \eqref{eq:curv:1} for $f^a$ transforms under a holomorphic gauge transformation as follows,
\begin{align}
 &f^a_{z\bar z} \mapsto g_a\,f^a_{z\bar z}\, g_a^{-1} = - \frac{nk}{N(1+z\bar z)^2} \mathbb{I}_m + 2\epsi\,\pd_{z}\pd_{\bar z} (g_a\, M^a g_a^{-1}), \label{eq:transformed:curvature:a} 
\end{align} 
where we have used that $g_a$ is constant. Since $g_a\,f^a g_a^{-1}$ must satisfy \eqref{eq:metric:curvature:a} with $h^a$ replaced by $(g_a^{-1})^{\dgr}h^a\,g_a^{-1}$, we can conclude from \eqref{eq:transformed:curvature:a} that  
\begin{align}
 (g_a^{-1})^{\dgr}h^a\,g_a^{-1} = (1+z\bar z)^{\frac{nk}{N}}\, e^{-2\epsi \, g_a M^a g_a^{-1}}.
\end{align}
This is consistent with \eqref{eq:herm:structure:a}, written in the form
\begin{align}
 (g_a^{-1})^{\dgr}h^a\,g_a^{-1} = (1+z\bar z)^{\frac{nk}{N}}\,(g_a^{-1})^{\dgr} e^{-2\epsi M^a} g_a^{-1}
\end{align}
if $(g_a^{-1})^{\dgr} = g_a$, i.e.~$g_a\in\U{m}$. Hence, in the regime of dissolving vortices, the transformation laws \eqref{eq:metric:a:trafo} and \eqref{eq:metric:b:trafo} are replaced by
\begin{align}
 &M^a \mapsto g_a M^a g_a^{-1}, \\
 &M^b \mapsto g_b M^b g_b^{-1},
\end{align}
where $g_a\in\U{m}$ and $g_b\in\U{n}$.

The matrices $W_r$ in \eqref{eq:W:Bog:1}, \eqref{eq:W:Bog:2} are acted upon by gauge transformations in the same way as $\phi$, i.e.
\begin{align}
 W_r \mapsto g_b W_r g_a^{-1}, \label{eq:W:unitary:action}
\end{align}
for all $r=0,\dots,k$. It follows that the moduli space of dissolving non-abelian vortices is the quotient
\begin{align}
 \curlM = \frac{ \lbrace (W_r)_{r=0,\dots,k} \in (\Mat{n}{m}{\CC})^{\oplus k+1} \colon \text{\eqref{eq:W:Bog:1} and \eqref{eq:W:Bog:2} hold} \rbrace }{ {\rm S}(\U{m}\times\U{n}) }. \label{eq:dissolving:moduli}
\end{align}
Strictly speaking we should quotient by ${\rm S}(\U{m}\times\U{n})/\ze_N$, but since the central subgroup $\ze_N$ acts trivially on the $W_r$, it is omitted.

\section{K\"ahler quotient construction of the moduli space} \label{sec:symplectic:quotient}

Vortex moduli spaces are generally K\"ahler \cite{GP:direct, GP:1994, Samols}, and the standard argument to show this is to give a construction of the moduli space as a K\"ahler quotient, see e.g.~\cite{Hitchin:et:al}. It is worthwhile to carry out the K\"ahler quotient construction of \eqref{eq:dissolving:moduli} explicitly since this immediately yields a natural metric on $\curlM$.  Moreover, we will see that this metric coincides with the metric derived from the kinetic energy of the fields near the Bradlow limit. Geodesics on the moduli space with respect to the latter metric determine the slow motion of vortices \cite{Manton:BPS:1982, Stuart:dynamics}. 

To identify  $\curlM$ in \eqref{eq:dissolving:moduli} as a K\"ahler quotient, we proceed in two steps: (i) We show that the space $(\Mat{n}{m}{\CC})^{\oplus k+1}$ is K\"ahler and (ii) that equations \eqref{eq:W:Bog:1}, \eqref{eq:W:Bog:2} define the level set of a moment map on $(\Mat{n}{m}{\CC})^{\oplus k+1}$. The K\"ahler quotient construction is a special case of the Marsden--Weinstein reduction, for which we refer to standard textbooks on symplectic geometry, e.g.~\cite{Silva, Berndt}.

The space $(\Mat{n}{m}{\CC})^{\oplus k+1}$ carries a hermitian metric, namely
\begin{align}
 H_{n\times m}^{(k+1)}((M_r)_r,(N_r)_r) = \sum_{r=0}^k \tr{ M_r N_r^{\dgr}}, \label{eq:hermitian:h}
\end{align}
for $(M_r)_r,(N_r)_r\in(\Mat{n}{m}{\CC})^{\oplus k+1}$. The corresponding Riemannian metric and K\"ahler form on $(\Mat{n}{m}{\CC})^{\oplus k+1}$ are 
\begin{align}
 g_{n\times m}^{(k+1)}((M_r)_r,(N_r)_r) &= \Re{H((M_r)_r,(N_r)_r)} \\
 					&= \half \sum_{r=0}^k \tr{M_r N_r^{\dgr} + N_r M_r^{\dgr}} , \\
 \omega_{n\times m}^{(k+1)}((M_r)_r,(N_r)_r) &= -\Im{H((M_r)_r,(N_r)_r)} \\ 
 					     &= \ihalf \sum_{r=0}^k \tr{M_r N_r^{\dgr} - N_r M_r^{\dgr}} .
\end{align}
Note that $\omega_{n\times m}^{(k+1)}$ is closed, which becomes apparent if we write 
\begin{align}
 \omega_{n\times m}^{(k+1)} = \im \sum_{r=0}^k \tr{\d{M_r}\w\d{M_r^{\dgr}}}.
\end{align}
Hence $(\Mat{n}{m}{\CC})^{\oplus k+1}$ is K\"ahler with the compatible complex structure being the standard one, i.e.~$(M_r)_r\mapsto (\im M_r)_r$.  

By \eqref{eq:W:unitary:action} the group $\U{m}\times\U{n}$ acts on $(\Mat{n}{m}{\CC})^{\oplus k+1}$, and the K\"ahler form $\omega_{n\times m}^{(k+1)}$ is invariant under this action. In the following subsection we give the moment map for this action, which suffices to show that \eqref{eq:dissolving:moduli} is a K\"ahler quotient.


\subsection{Moment maps} \label{sec:mom:maps:symp:quotient}

We first identify the moment map for the $\U{m}$-action on $(\Mat{n}{m}{\CC})^{\oplus k+1}$. Let $X\in\u{m}$, i.e.~$X$ is an anti-hermitian  $m\!\times\!m$ matrix. Through the $\U{m}$-action, $X$ generates a flow on $(\Mat{n}{m}{\CC})^{\oplus k+1}$,
\begin{align}
 (M_r)_r \mapsto (M_r\e^{-tX})_r = (M_r - t M_r X)_r,
\end{align}
up to linear order in $t\in\RR$. Thus $X$ induces a vector field on $(\Mat{n}{m}{\CC})^{\oplus k+1}$, namely
\begin{align}
 \tilde X_{(M_r)_r} = (-M_r X)_r.
\end{align}
The subscript on the left-hand side indicates the base point in $(\Mat{n}{m}{\CC})^{\oplus k+1}$ at which the field $\tilde X$ is evaluated.
The map 
\begin{align}
 \mom_m \colon(\Mat{n}{m}{\CC})^{\oplus k+1} &\to \u{m} \\
				    (M_r)_r &\mapsto \ihalf \sum_{r=0}^k M_r^{\dgr}M_r
\end{align}
can be checked to satisfy the defining equation of a moment map, namely
\begin{align}
 \langle X, \d\mom_m((N_r)_r) \rangle = \omega_{n\times m}^{(k+1)}(\tilde{X}_{(M_r)_r}, (N_r)_r), \label{eq:mom:map:m}
\end{align}
where angle brackets denote the pairing $\langle X, Y\rangle = - \tr{XY}$ for $X,Y\in\u{m}$. To see that \eqref{eq:mom:map:m} holds, one only needs to note that 
\begin{align}
 \d\mom_m((N_r)_r) = \ihalf \sum_{r=0}^k (N_r^{\dgr}M_r + M_r^{\dgr}N_r).
\end{align}
We are free to add a term to $\mom_m$ that is proportional to $\im\mathbb{I}_m$ without spoiling \eqref{eq:mom:map:m}. Therefore, taking
\begin{align}
 \mom_m((M_r)_r) = \ihalf \sum_{r=0}^k M_r^{\dgr}M_r - \im \frac{2n}{N} \mathbb{I}_m
\end{align}
as our final version of the moment map for the $\U{m}$-action, equation \eqref{eq:W:Bog:1} can be written as
\begin{align}
 \mom_m((W_r)_r) = 0, \quad (W_r)_r\in(\Mat{n}{m}{\CC})^{\oplus k+1}.
\end{align}
This equation is clearly preserved by the $\U{m}$-action.

The moment map for the $\U{n}$-action is obtained analogously. An element $X\in\u{n}$ generates a vector field
\begin{align}
 \tilde X_{(M_r)_r} = (XM_r )_r,
\end{align}
for $(M_r)_r\in(\Mat{n}{m}{\CC})^{\oplus k+1}$. Then
\begin{align}
 \mom_n \colon(\Mat{n}{m}{\CC})^{\oplus k+1} &\to \u{n} \\
				    (M_r)_r &\mapsto -\ihalf \sum_{r=0}^k M_rM_r^{\dgr} +  \im \frac{2m}{N}\mathbb{I}_n
\end{align}
satisfies
\begin{align}
 \langle X, \d\mom_n((N_r)_r) \rangle = \omega_{n\times m}^{(k+1)}(\tilde{X}_{(M_r)_r}, (N_r)_r). 
\end{align}
Here angle bracktes are the analogous pairing in $\u{n}$, i.e.~$\langle X, Y\rangle = - \tr{XY}$ for $X,Y\in\u{n}$.
Equation \eqref{eq:W:Bog:2} becomes
\begin{align}
 \mom_n((W_r)_r) = 0, \quad (W_r)_r\in(\Mat{n}{m}{\CC})^{\oplus k+1},
\end{align}
which is left invariant by the $\U{n}$-action.

Now we have the K\"ahler quotient
\begin{align}
 \curlM' = \frac{\mom_m^{-1}(0)\cap\mom_n^{-1}(0)}{\U{m}\times\U{n}}.
\end{align}
The K\"ahler form on $\curlM'$ is inherited from the restriction of $\omega_{n\times m}^{(k+1)}$ to $\mom_m^{-1}(0)\cap\mom_n^{-1}(0)$ \cite{Hitchin:et:al}. In order to identify $\curlM'$ with the quotient in \eqref{eq:dissolving:moduli}, we need the observation that the overall determinant of
\begin{align}
 \left( \begin{array}{cc} g_a & 0 \\ 0 & g_b \end{array} \right) \in \U{m}\times\U{n}
\end{align}
acts trivially on $(\Mat{n}{m}{\CC})^{\oplus k+1}$. Therefore,
\begin{align}
 \curlM' = \frac{\mom_m^{-1}(0)\cap\mom_n^{-1}(0)}{{\rm S}(\U{m}\times\U{n})} = \curlM.
\end{align}

\subsection{The moduli space metric} \label{sec:moduli:space:metric}

The moduli space $\curlM$ carries a physically motivated metric. So far we have only considered the static Yang--Mills--Higgs energy functional \eqref{eq:energy}, and static field configurations $a$, $b$, $\phi$. One can make the fields time-dependent, but then \eqref{eq:energy} must be augmented by a kinetic energy term,
\begin{align}
 T = \quart \int_{\CP{1}} \tr{\dot{\phi}^{\dgr}\dot{\phi}} \dvol_{\CP{1}} -  \half \int_{\CP{1}}  \tr{\dot{a}\w *\dot{a}^{\dgr}} -  \half \int_{\CP{1}}  \tr{\dot{b}\w *\dot{b}^{\dgr}} , \label{eq:kinetic:energy}
\end{align}
where $\dot{}$ indicates differentiation with respect to time. Note that $T$ is written in temporal gauge, where the time components of the gauge fields are set to zero. It is assumed that for slow changes in time the fields $a$, $b$, $\phi$ depend on time only through their moduli. The triple $(\dot{a}, \dot{b}, \dot{\phi})$ can be regarded as a tangent vector to the moduli space. Then $T$ defines a metric on the moduli space. Geodesics with respect to this metric describe slow vortex motion, as was first explained in the context of monopoles in \cite{Manton:BPS:1982} and proven for abelian vortices in \cite{Stuart:dynamics}.

We analyze $T$ near the Bradlow limit, i.e.~we evaluate \eqref{eq:kinetic:energy} on solutions of \eqref{eq:W:Bog:1}, \eqref{eq:W:Bog:2}, instead of solutions of the full Bogomolny equations. Consistency then requires that we only keep contributions to $T$ which are of first order in $\epsi$.

Recall that \eqref{eq:W:Bog:1}, \eqref{eq:W:Bog:2} were derived in holomorphic gauge. Since \eqref{eq:kinetic:energy} defines $T$ in unitary gauge, solutions of \eqref{eq:W:Bog:1}, \eqref{eq:W:Bog:2} first have to be brought into unitary gauge, where $h^a=\mathbb{I}_m$ and $h^b=\mathbb{I}_n$. This is achieved by smooth maps $g_a\colon U_0\to\GLi{m}$, $g_b\colon U_0\to\GLi{n}$ such that
\begin{align}
 &h^a(z, \bar z) = (1+z\bar z)^{\frac{nk}{N}}\,e^{-2\epsi M^a(z, \bar z)} = g_a(z, \bar z)^{\dgr} g_a(z, \bar z), \\
 &h^b(z, \bar z) = (1+z\bar z)^{-\frac{mk}{N}}\,\,e^{-2\epsi M^b(z, \bar z)} = g_b(z, \bar z)^{\dgr} g_b(z, \bar z).
\end{align}
Up to first order in $\epsi$,
\begin{align}
 &g_a(z, \bar z) = (1+z\bar z)^{\frac{nk}{2N}}\,(\mathbb{I}_m-\epsi M^a(z, \bar z)), \\
 &g_b(z, \bar z) = (1+z\bar z)^{-\frac{mk}{2N}}\,(\mathbb{I}_n-\epsi M^b(z, \bar z)). 
\end{align}
Since $\phi \mapsto g_b\phi g_a^{-1}$, we have in unitary gauge and up to lowest order in $\epsi$, 
\begin{align}
 \phi(z, \bar z) = \sqrt{\epsi}\,(1+z\bar z)^{-\frac{k}{2}}\sum_{r=0}^k V_{r} z^r. 
\end{align}
Bringing the gauge field into unitary gauge, we obtain
\begin{align}
	A_z = \left(\begin{array}{cc} a_z & 0 \\ 0 & b_z \end{array}\right)
	&= \left( \begin{array}{cc} (h^a)^{-1}\pd_z h^a + g_a (\pd_z g_a^{-1}) & 0 \\ 0 & (h^b)^{-1}\pd_z h^b + g_b (\pd_z g_b^{-1}) \end{array}\right) \\
	&= \left( \begin{array}{cc} \frac{nk}{2N}\frac{\bar z}{1+z\bar z} - \epsi\,\pd_z M^a & 0 \\ 0 & \frac{mk}{2N}\frac{\bar z}{1+z\bar z} - \epsi\,\pd_z M^b \end{array}\right),
\end{align}
and
\begin{align}
	A_{\bar z} = \left(\begin{array}{cc} a_{\bar z} & 0 \\ 0 & b_{\bar z} \end{array}\right)
	&= \left( \begin{array}{cc} g_a (\pd_{\bar z} g_a^{-1}) & 0 \\ 0 & g_b (\pd_{\bar z} g_b^{-1}) \end{array}\right) \\
	&= \left( \begin{array}{cc} -\frac{nk}{2N}\frac{z}{1+z\bar z} + \epsi\,\pd_{\bar z} M^a & 0 \\ 0 & -\frac{mk}{2N}\frac{z}{1+z\bar z} + \epsi\,\pd_{\bar z} M^b \end{array}\right), 
\end{align}

We now substitute the above expressions for $\phi$, $a$, and $b$ into \eqref{eq:kinetic:energy}. From the previous formulae it follows that $\dot{a}$ and $\dot{b}$ are of order $\epsi$, and hence the second and third integral in \eqref{eq:kinetic:energy} are of order $\epsi^2$ and can be neglected. Thus
\begin{align}
 T &= \quart \int_{\CP{1}} \tr{\dot{\phi}^{\dgr}\dot{\phi}} \dvol_{\CP{1}} \\
   &= \frac{\epsi}{4} \tr{\dot{V}_r^{\dgr} \dot{V}_{s}} \int  \frac{\bar{z}^r z^s}{(1+z\bar z)^k} \ihalf \frac{4R^2}{(1+z\bar z)^2} \d{z}\w\d{\bar z} \\
   &= \im R^2 \frac{\epsi}{2} \tr{\dot{V}_r^{\dgr} \dot{V}_{s}} \underbrace{\int  \frac{\bar{z}^r z^s}{(1+z\bar z)^{k+2}} \d{z}\w\d{\bar z}}_{(-2\pi\im)\de_{r,s} I_{r,k}} \\
   &= \frac{\epsi}{4} \vol{\CP{1}}\, \tr{\dot{W}_r^{\dgr} \dot{W}_{r}},
\end{align}
where we used $\vol{\CP{1}} = 4\pi R^2$, and summation over the repeated indices $r, s$ is implied. Since $\vol{\CP{1}} = 4\pi R_B^2(1+\epsi)^2$, the above expression for $T$ contains contributions which are of second and third order in $\epsi$. Neglecting those terms and using $R_B^2 = k$,
\begin{align}
 T &= \pi k \epsi\, \tr{\dot{W}_r^{\dgr} \dot{W}_{r}}.
\end{align}
In order to turn this expression for $T$ into a metric on the moduli space $\curlM$, the constraints \eqref{eq:W:Bog:1}, \eqref{eq:W:Bog:2} must be accounted for and a gauge fixing scheme must be applied. We will do this explicitly in a special case in section \ref{sec:Grass:applications}. For now, observe that $T$ essentially agrees with the Riemannian metric on $(\Mat{n}{m}{\CC})^{\oplus k+1}$,
\begin{align}
 T &= \pi k \epsi\, g_{n\times m}^{(k+1)}((\dot{W}_r)_r, (\dot{W}_{r})_r). \label{eq:kinetic:energy:h}
\end{align}
From this it follows that the physically motivated metric on the moduli space $\curlM$, inherited from the kinetic energy $T$, agrees with the metric that $\curlM$ inherits through the K\"ahler quotient construction, up to an overall constant factor. This is generally true, not only near the Bradlow limit, as we briefly explain in the next paragraph. 

The kinetic energy \eqref{eq:kinetic:energy} is the standard $L^2$-metric on the tangent space of ${\mathfrak A}(P)\times\Gamma(\Sigma,E_{21})$. Here ${\mathfrak A}(P)$ denotes the space of unitary connections on $P$, i.e.~connections given by $\covd = \d + A$ with $A^{\dgr} = -A$, and $\Gamma(\Sigma,E_{21})$ is the space of smooth sections of the vector bundle $E_{21}$. By a K\"ahler quotient construction analogous to \cite{GP:1994} the moduli space is a subquotient of ${\mathfrak A}(P)\times\Gamma(\Sigma,E_{21})$, and hence its metric is inherited from \eqref{eq:kinetic:energy}. 

The K\"ahler quotient construction also leads to an interesting observation regarding the complex structure of the moduli space: The space of fields ${\mathfrak A}(P)\times\Gamma(\Sigma,E_{21})$ is K\"ahler with the compatible complex structure being
\begin{align}
 I \colon 
 (\dot{a},\dot{b},\dot{\phi}) &\mapsto (*\dot{a},*\dot{b},\im\dot{\phi}),
\end{align}
where $(\dot{a},\dot{b},\dot{\phi})\in T_{(a,b,\phi)}\!\left({\mathfrak A}(P)\times\Gamma(\Sigma,E_{21})\right)$ is a tangent vector at $(a,b,\phi)\in{\mathfrak A}(P)\times\Gamma(\Sigma,E_{21})$. The gauge fixing condition that must be imposed on \eqref{eq:kinetic:energy} in order to obtain a well-defined metric on the moduli space is given by the equations
\begin{align}
 &*(\d*\dot{a} + *\dot{a}\w a + a\w*\dot{a}) - \quart (\phi^{\dgr}\dot{\phi} - \dot{\phi}^{\dgr}\phi) = 0, \label{eq:Gauss:1} \\
 &*(\d*\dot{b} + *\dot{b}\w b + b\w*\dot{b}) + \quart (\dot{\phi}\phi^{\dgr} - \phi\dot{\phi}^{\dgr}) = 0. \label{eq:Gauss:2}
\end{align} 
These are a direct generalization of Gauss' law for abelian vortices (cf.~\cite{Manton:Sutcliffe, Samols}). Equations \eqref{eq:Gauss:1}, \eqref{eq:Gauss:2} ensure that $I$ descends to a complex structure on the moduli space, rendering the moduli space K\"ahler. We have thus given an explicit example of the abstract argument in \cite{Hitchin:et:al}.

\section{Moduli spaces and their dimensions} \label{sec:dimensions:examples}

The dimension of the moduli space of vortices is a function of the three integral parameters $m$, $n$, and $k$. In this section we obtain the dimensions of the moduli space $\curlM$ of dissolving vortices for different combinations of $m$, $n$, and $k$. The dimension of $\curlM$ agrees with the dimension of the moduli space of general vortices on $\CP{1}$, not necessarily dissolving, because increasing the area of $\CP{1}$ is a smooth process. Note, however, that the general moduli space may have components of lower dimension. For example, the maximally abelian solutions studied in \cite{Manton:Rink:nonab} lie in such a component.

In the following we always assume $m\ge n$, which presents no loss of generality. Before looking at examples of moduli spaces for special combinations of the parameters $m$, $n$, and $k$, we use basic linear algebra to exclude a range of values of $m$ for which the moduli space is empty.

\subsection{The Bogomolny equations and linear maps}

The matrices $W_r$, $r=0,\dots,k$, can be regarded as linear maps,
\begin{align}
 &W_r \colon \CC^m \to \CC^n, \\
 &W_r^{\dgr}\colon \CC^n\to \CC^m.
\end{align}
We rewrite \eqref{eq:W:Bog:1} as
\begin{align}
 \left(\begin{array}{ccc} W_0^{\dgr} & \dots & W_k^{\dgr} \end{array}\right) \left(\begin{array}{c} W_0 \\ \vdots \\ W_k \end{array}\right) 
 = \frac{4n}{N} \mathbb{I}_m.
\end{align}
The second matrix on the left-hand side represents a linear map $\CC^{m}\to\CC^{(k+1)n}$. For $m>(k+1)n$ the kernel of this map is non-trivial. Hence there exists a non-zero $v\in\CC^m$ such that
\begin{align}
 \left(\begin{array}{ccc} W_0^{\dgr} & \dots & W_k^{\dgr} \end{array}\right) \underbrace{\left(\begin{array}{c} W_0 \\ \vdots \\ W_k \end{array}\right)v}_{=0} 
 \ne \frac{4n}{N} v.
\end{align}
Therefore \eqref{eq:W:Bog:1} cannot have solutions for $m>(k+1)n$, and thus the moduli space $\curlM$ is empty.

In the previous argument we did not make use of \eqref{eq:W:Bog:2}. An analogous analysis can be carried out starting with \eqref{eq:W:Bog:2} in the form
\begin{align}
 \left(\begin{array}{ccc} W_0 & \dots & W_k \end{array}\right) \left(\begin{array}{c} W_0^{\dgr} \\ \vdots \\ W_k^{\dgr} \end{array}\right) 
 = \frac{4m}{N} \mathbb{I}_n.
\end{align}
This time the second matrix on the left-hand side is a linear map $\CC^{n}\to\CC^{(k+1)m}$. For $n>(k+1)m$ there exists a non-zero $v\in\CC^n$ such that
\begin{align}
 \left(\begin{array}{ccc} W_0 & \dots & W_k \end{array}\right) \underbrace{\left(\begin{array}{c} W_0^{\dgr} \\ \vdots \\ W_k^{\dgr} \end{array}\right)v}_{=0}  \ne \frac{4m}{N} v.
\end{align}
However, since we have assumed $m\ge n$, the situation $n>(k+1)m$ does not occur.

In the following we will be interested in non-empty examples of moduli spaces. Therefore we always assume that $m\le(k+1)n$.

\subsection{The case $n=1$ and Grassmannians} \label{sec:n:1:Grassmann}

Taking $n=1$ renders \eqref{eq:W:Bog:2} redundant since it is simply the trace of \eqref{eq:W:Bog:1}. Thus we are left with
\begin{align}
 \left(\begin{array}{ccc} W_0^{\dgr} & \dots & W_k^{\dgr} \end{array}\right) \left(\begin{array}{c} W_0 \\ \vdots \\ W_k \end{array}\right) 
 = \frac{4}{N} \mathbb{I}_m. \label{eq:W:unitary:basis}
\end{align}
The gauge group that acts on $(\Mat{1}{m}{\CC})^{\oplus k+1}$ is ${\rm S}(\U{m}\times\U{1})$. Elements of ${\rm S}(\U{m}\times\U{1})$ can be written as
\begin{align}
	\left(\begin{array}{cc} g_a & 0 \\ 0 & \det{g_a}^{-1} \end{array}\right), \quad g_a\in\U{m},
\end{align}
and their action on $(W_r)_{r=0,\dots,k}$ is given by
\begin{align}
 \left(\begin{array}{c} W_0 \\ \vdots \\ W_k \end{array}\right) 
 \mapsto \left(\begin{array}{c} W_0 \\ \vdots \\ W_k \end{array}\right) {\tilde g}^{-1} 
 = \left(\begin{array}{c} W_0 {\tilde g}^{-1} \\ \vdots \\ W_k {\tilde g}^{-1} \end{array}\right),
\end{align}
where ${\tilde g} = \det{g_a}g_a \in\U{m}$. Therefore
\begin{align}
 \curlM = \frac{\mom_m^{-1}(0)\cap\mom_1^{-1}(0)}{{\rm S}(\U{m}\times\U{1})} = \frac{\mom_m^{-1}(0)}{\U{m}} = \Grass{m}{k+1}.
\end{align}
The last equality is the definition of the Grassmannian as a K\"ahler quotient (see appendix \ref{sec:Grass:quotient} for details). A more direct way to see that the moduli space is a Grassmannian is as follows: Since $n=1$, the $W_r$ are row vectors with $m$ entries. Equation \eqref{eq:W:unitary:basis} says that the $m$ columns of the matrix
\begin{align}
 \left(\begin{array}{c} W_0 \\ \vdots \\ W_k \end{array}\right) 
\end{align}
form a unitary basis for an $m$-dimensional subspace $\Lambda\subset\CC^{k+1}$. The right-action by $\U{m}$ transforms one unitary basis of $\Lambda$ into another. Hence $\U{m}$-orbits of solutions to \eqref{eq:W:unitary:basis} are in $1:1$ correspondence with points in $\Grass{m}{k+1}$.

We remark on two special cases: The first case is $m=1$, which leads to
\begin{align}
	\Grass{1}{k+1} = \CP{k}.
\end{align}	
This is the moduli space of abelian vortices on $\CP{1}$, and its properties near the Bradlow limit were studied in \cite{Baptista:Manton:S2}. The second case is $m=k+1$. Now we have 
\begin{align}
	\Grass{k+1}{k+1} = \rm{pt.},
\end{align}
i.e.~the moduli space degenerates to a single point. This behaviour was to be expected based on the earlier observation that the moduli space is empty for $m>k+1$. The fact that the moduli space degenerates to a point shows that there are non-abelian vortices which are extended objects, unlike abelian vortices. Moreover, $\curlM = \rm{pt.}$ implies that a configuration with vortex number $k$ and $m=k+1$ extends over the entire $\CP{1}$. This is an example of a vortex configuration where individual vortices cannot be localized, as promised in the introduction of this paper.

We will study the properties of the moduli spaces for $n=1$ in more detail in section \ref{sec:Grass:applications}. The following table summarizes the moduli spaces for $n=1$ and fixed $k$, and their dimensions. 
\begin{align}
 \begin{array}{l|ccc}  m= & 1 & m & k+1 \\ \hline \text{space:} & \CP{k} & \Grass{m}{k+1} & \rm{pt.} \\ \text{dim:} & k & m(k+1-m) & 0 \end{array} \nonumber
\end{align}

\subsection{The moduli space for $m=(k+1)n$} \label{sec:m:k+1:n}

We claim that generally, when $m=(k+1)n$, the moduli space is a point. This provides another class of non-abelian vortices that stretch out over the entire $\CP{1}$. 

To see that the moduli space is a point, consider again \eqref{eq:W:Bog:1} in the form
\begin{align}
	Z^{\dgr} Z = \frac{4n}{N} \mathbb{I}_{(k+1)n},
\end{align}
where
\begin{align}
	Z =  \left(\begin{array}{c} W_0 \\ \vdots \\ W_k \end{array}\right) \colon \CC^{(k+1)n} \to \CC^{(k+1)n}.
\end{align}
Thus $Z$ is essentially a unitary matrix. The gauge group $\U{m}=\U{(k+1)n}$ acts on $Z$ on the right, as in the previous subsection. We can use this gauge freedom to perform the transformation
\begin{align}
	Z \mapsto Z g^{-1} = 2\sqrt{\frac{n}{N}}\, \mathbb{I}_{(k+1)n},
\end{align}
where $g\in\U{(k+1)n}$. Hence all solutions of \eqref{eq:W:Bog:1} lie in the same $\U{(k+1)n}$-orbit. This proves our claim.

The above expression for $Z g^{-1}$ implies
\begin{align}
	W_r g^{-1} = 2\sqrt{\frac{n}{N}} \left(\begin{array}{c|c|c|c|c|c|c}  0 & \cdots & 0 & \mathbb{I}_n & 0 & \cdots & 0 \end{array}\right),
\end{align}
where the unit matrix sits in the $r$-th block of size $n\!\times\!n$. We use this to verify \eqref{eq:W:Bog:2},
\begin{align}
	\frac{1}{4} W_r W_r^{\dgr} =  \frac{n}{N} (k+1) \mathbb{I}_n.
\end{align}

\subsection{Dimension of the moduli space for $m=n$} \label{sec:m:n}

For $m=n$ the Bogomolny equations for dissolving vortices \eqref{eq:W:Bog:1}, \eqref{eq:W:Bog:2} have the symmetry $W_r \mapsto W_r^{\dgr}$, $r=0,\dots,k$, which should yield an involution on the moduli space. Here we consider two special solutions of \eqref{eq:W:Bog:1}, \eqref{eq:W:Bog:2} with $m=n$. By studying small perturbations of these solutions we obtain the dimension of the tangent space of $\curlM$. The Bogomolny equations for non-abelian vortices with $m=n$ were studied in \cite{Manton:Sakai} on a general Riemann surface. 

First consider
\begin{align}
	&W_0 = 2\sqrt{\frac{m}{N}}\, \mathbb{I}_m, \\
	&W_r = 0, \quad\text{for }r=1,\dots,k.
\end{align}
This clearly is a solution of \eqref{eq:W:Bog:1}, \eqref{eq:W:Bog:2}. Now introduce small perturbations $\de_0,\de_r\in\Mat{m}{m}{\CC}$, $r=1,\dots,k$,
\begin{align}
	&W_0 = 2\sqrt{\frac{m}{N}}\, \mathbb{I}_m + \de_0, \\
	&W_r = \de_r.
\end{align}
Writing down \eqref{eq:W:Bog:1}, \eqref{eq:W:Bog:2} and keeping terms only up to linear order in $\de_0,\de_r$, we obtain the constraint
\begin{align}
	\de_0^{\dgr} + \de_0 = 0.
\end{align}
Thus we have the following numbers of real degrees of freedom,
\begin{align}
	&\de_0 \colon\: m^2, \\
	&\de_r \colon\: 2m^2,
\end{align}
i.e.~$(2k+1)m^2$ real degrees of freedom in total. To remove the gauge degrees of freedom, we consider the infinitesimal action of $\U{m}\!\times\!\U{m}$ on our special solution. Let $g_a, g_b\in\U{m}$, and expand
\begin{align}
	&g_a = \mathbb{I}_m + X_a, \\
	&g_b = \mathbb{I}_m + X_b, 
\end{align}
with matrices $X_a, X_b\in\u{m}$. Then
\begin{align}
	&W_0 = 2\sqrt{\frac{m}{N}}\, \mathbb{I}_m \mapsto g_b W_0 g_a^{-1} = 2\sqrt{\frac{m}{N}}\, (\mathbb{I}_m + X_b - X_a), \\
	&W_r = 0 \mapsto 0, \quad\text{for }r=1,\dots,k.
\end{align}
The combination $X_b - X_a$ is an anti-hermitian matrix, carrying $m^2$ real degrees of freedom. These gauge degrees of freedom can be used to cancel the perturbation $\de_0$. Therefore the real dimension of the tangent space of $\curlM$ at our special solution is $2km^2$. Hence
\begin{align}
	\dim{\curlM} = km^2,
\end{align}
by which is meant the complex dimension of $\curlM$. 

We are led to the same result for the dimension of $\curlM$ if we start with a different, more symmetric solution of \eqref{eq:W:Bog:1}, \eqref{eq:W:Bog:2},
\begin{align}
	W_r = 2\sqrt{\frac{1}{k+1}\frac{m}{N}}\, \mathbb{I}_m, 
\end{align}
for all $r=0,\dots,k$. Plugging the perturbed solution $W_r+\de_r$ into \eqref{eq:W:Bog:1}, \eqref{eq:W:Bog:2} yields the linear constraint
\begin{align}
	\sum_{r=0}^k (\de_r^{\dgr} + \de_r) = 0.
\end{align}
This constraint reduces the number of real degrees of freedom in the $\de_r$ from $2(k+1)m^2$ to $(2k+1)m^2$. Acting on the $W_r$ with the same infinitesimal gauge transformations as above yields
\begin{align}
	g_b W_r g_a^{-1} = 2\sqrt{\frac{1}{k+1}\frac{m}{N}}\, (\mathbb{I}_m + X_b - X_a).
\end{align}
Hence the anti-hermitian matrix $X_b - X_a$ allows for $m^2$ real degrees of freedom to be removed from the $\de_r$. Altogether we find again that $2km^2$ real degrees of freedom remain.

Note that in our analysis of the gauge degrees of freedom we have implicitly used that
\begin{align}
	\curlM = \frac{\mom_m^{-1}(0)\cap\mom_m^{-1}(0)}{\U{m}\times\U{m}}, 
\end{align}
which was established at the end of subsection \ref{sec:mom:maps:symp:quotient}. Had we taken only gauge transformations with overall determinant one, the infinitesimal gauge transformations would have been required to satisfy
\begin{align}
	\tr{X_a} + \tr{X_b} = 0.
\end{align}
Since this puts no constraint on the trace of $X_b - X_a$, our counting arguments in this subsection are unaffected.

\subsection{The case $k=0$}

We saw that for $k=0$ the Bogomolny equations reduce to \eqref{eq:V0:Bog:1}, \eqref{eq:V0:Bog:2} near the Bradlow limit. If $m$ is strictly greater than $n$, the linear map
\begin{align}
	V_0 \colon \CC^m \to \CC^n
\end{align}
has a non-zero $v\in\CC^m$ in its kernel. By \eqref{eq:V0:Bog:1},
\begin{align}
	\frac{1}{2} V_0^{\dgr} \underbrace{V_0 v}_{=0} \ne \frac{n}{N} v,
\end{align}
Hence the moduli space is empty for $m>n$.

For $m=n$ equations \eqref{eq:V0:Bog:1}, \eqref{eq:V0:Bog:2} say that $V_0$ is essentially a unitary matrix. We can use either the left-action or the right-action of $\U{m}$ to set
\begin{align}
	V_0 = \sqrt{\frac{2m}{N}}\, \mathbb{I}_m.
\end{align}
Hence $\curlM = \rm{pt.}$, which is consistent with subsections \ref{sec:m:k+1:n} and \ref{sec:m:n} setting $k=0$.

\section{Grassmannians and applications} \label{sec:Grass:applications}

We now look more closely at the moduli space of non-abelian vortices for $n=1$. We saw in the previous section that the moduli space is the Grassmannian $\Grass{m}{k+1}$. In the present section we first identify the moduli space metric from subsection \ref{sec:moduli:space:metric} with a standard metric on $\Grass{m}{k+1}$ that generalizes the Fubini--Study metric on complex projective space. We then use this result to calculate the volume of the moduli space and study the statistical mechanics of dissolving vortices.

\subsection{The moduli space metric revisited} \label{sec:Grass:moduli:metric}

We start with writing expression \eqref{eq:kinetic:energy:h} for the kinetic energy as
\begin{align}
 T = \pi k \epsi\, H_{1\times m}^{(k+1)}((\dot{W}_r)_r, (\dot{W}_{r})_r).
\end{align}
Let us introduce the matrix $Z$,
\begin{align}
 Z = \left(\begin{array}{c} W_0 \\ \vdots \\ W_k \end{array}\right) \in\Mat{(k+1)}{m}{\CC},
\end{align}
and the hermitian metric
\begin{align}
 \langle Z, Z \rangle = \tr{ Z Z^{\dgr}} = H_{1\times m}^{(k+1)}((W_r)_r, (W_r)_r).
\end{align}
Then the kinetic energy reads $T = \pi k \epsi\, \langle \dot{Z}, \dot{Z} \rangle$, and we wish to identify the metric that $\langle\cdot,\cdot\rangle$ induces on the moduli space $\curlM=\Grass{m}{k+1}$.

Equation \eqref{eq:W:Bog:1} is written in terms of the new matrix $Z$ as
\begin{align}
	Z^{\dgr} Z = \frac{4}{m+1}\, \mathbb{I}_m. \label{eq:Z:Bog}
\end{align}
To gauge fix $\langle\cdot,\cdot\rangle$, we need to characterize tangent vectors $\dot{Z}$ that are pure gauge, i.e.~are generated by the $\U{m}$-action
\begin{align}
	Z \mapsto Z g^{-1}, \quad g\in\U{m}.
\end{align}
Let $\tau_i$ be a basis of the Lie algebra $\u{m}$ such that $\langle\tau_i,\tau_j\rangle = \de_{ij}$. To obtain the gauge fixed tangent vector $\dot{Z}_{\rm g.f.}$, we project onto directions in $\dot{Z}$ which are orthogonal to the directions generated by the $\tau_i$,
\begin{align}
 \dot{Z}_{\rm{g.f.}} 
 &= \dot{Z} - \sum_i \frac{\langle\dot{Z},Z\tau_i\rangle}{\langle Z\tau_i,Z\tau_i\rangle} Z\tau_i, \\
 & = \dot{Z} - \frac{1}{4}(m+1) \sum_i \langle\dot{Z},Z\tau_i\rangle Z\tau_i,
\end{align}
where we used \eqref{eq:Z:Bog} in going to the second line. We thus find
\begin{align}
	\langle \dot{Z}, \dot{Z} \rangle_{\rm{g.f.}} 
	&= \langle \dot{Z}_{\rm{g.f.}}, \dot{Z}_{\rm{g.f.}} \rangle \\
	&= \langle\dot{Z},\dot{Z}\rangle - \frac{1}{4}(m+1) \sum_i \langle\dot{Z},Z\tau_i\rangle \langle Z\tau_i,\dot{Z}\rangle, \\
	&= \tr{\dot{Z}\dot{Z}^{\dgr} \left(\mathbb{I}_{k+1} - \frac{1}{4}(m+1) ZZ^{\dgr}\right) }.
\end{align}
In the last line we used that the combination $Z^{\dgr}\dot{Z}$ is anti-hermitian, which follows by differentiating \eqref{eq:Z:Bog}, and hence
\begin{align}
	Z^{\dgr}\dot{Z} = \sum_i \langle Z^{\dgr}\dot{Z},\tau_i \rangle \tau_i.
\end{align}

If we introduce the new coordinate 
\begin{align}
	\curlT = \frac{1}{2}\sqrt{m+1}\, Z^{\dgr}
\end{align}
on the moduli space, equation \eqref{eq:Z:Bog} reads
\begin{align}
	\curlT\curlT^{\dgr} = \mathbb{I}_m, \label{eq:T:unitarity}
\end{align}
and the moduli space metric takes the form
\begin{align}
	\langle \dot{Z}, \dot{Z} \rangle_{\rm{g.f.}} 
	= H_{\curlM}(\dot{\curlT},\dot{\curlT}) 
	= \frac{4}{m+1} \tr{\dot{\curlT}\left(\mathbb{I}_{k+1} -  \curlT^{\dgr}\curlT\right)\dot{\curlT}^{\dgr} }.
\end{align}
From $H_{\curlM}$ we obtain the Riemannian metric $g_{\curlM}$ and K\"ahler form $\omega_{\curlM}$ on $\curlM$. It follows that
\begin{align}
T = \pi k \epsi\, g_{\curlM}(\dot{\curlT},\dot{\curlT}) , \label{eq:kin:energy:Riemannian}
\end{align}
and
\begin{align}
 \omega_{\curlM} = \im \frac{2}{m+1} \, \tr{\d\curlT \w \d\curlT^{\dgr} - \d\curlT\, \curlT^{\dgr} \w \curlT \d\curlT^{\dgr}}.
\end{align}
This is a scaled version of the Fubini--Study form $\omega_{\rm FS}$ on $\Grass{m}{k+1}$,
\begin{align}
 \omega_{\curlM} = \frac{4\pi}{m+1}\, \omega_{\rm FS}. \label{eq:Kahler:FS:form}
\end{align}
A careful derivation of $\omega_{\rm FS}$ is given in appendix \ref{sec:Grass:Fubini:Study}.

\subsection{Statistical mechanics on the moduli space} \label{sec:Grass:stat:mech}

Because of \eqref{eq:kin:energy:Riemannian} the motion of vortices follows geodesics on the moduli space with respect to the Riemannian metric $g_{\curlM}$. For non-abelian vortices we have no clear notion of individual vortices since vortices cannot generally be localized in small regions on $\CP{1}$, as we saw in section \ref{sec:dimensions:examples}. However, for large $k$ it is more meaningful to study statistical mechanics on the moduli space rather than the dynamics of individual vortices. This is because the dimension of $\Grass{m}{k+1}$, and hence the number of degrees of freedom grow with $k$. Statistical mechanics of abelian vortices has been studied before in \cite{Manton:stat:mech, Manton:Nasir}.

To allow $k$ to be large while remaining in the regime of dissolving vortices, $R_B$ has to increase accordingly. Recall that
\begin{align}
 R^2 = R_B^2(1+\epsi)^2 = k (1+\epsi)^2.
\end{align}
Hence the ratio $R^2/k$ should be kept fixed. For brevity we denote the complex dimension of the moduli space $\curlM = \Grass{m}{k+1}$ as $d$, i.e.
\begin{align}
 d = m(k+1-m).
\end{align}
Thinking of each fibre of the tangent bundle $T\curlM$ as a real vector space, we introduce the basis
\begin{align}
 &Q_{2i-1}, \\
 &Q_{2i} = \im Q_{2i-1},
\end{align}
where $i=1,\dots,d$. Then the tangent vector $\dot{\curlT}$ can be expanded as
\begin{align}
 \dot{\curlT} = \sum_{i=1}^{2d}\dot{\curlT}^i Q_i.  
\end{align}
Defining
\begin{align}
 g_{ij} = g_{\curlM}(Q_i,Q_j), 
\end{align}
we can express \eqref{eq:kin:energy:Riemannian} as
\begin{align}
 T = \pi k \epsi \sum_{i,j=0}^{2d} g_{ij}\dot{\curlT}^i\dot{\curlT}^j.
\end{align}
In the remainder of this section we use the convention that repeated upstairs and downstairs indices are summed over.

We take as the Lagrangian $L$ for the dynamics on the moduli space the kinetic energy $T$, i.e.
\begin{align}
 L = T = \pi k \epsi\, g_{ij}\dot{\curlT}^i\dot{\curlT}^j.  
\end{align}
With the conjugate momenta $\curlP_i$,
\begin{align}
 \curlP_i = \frac{\pd L}{\pd\dot{\curlT}^i} = 2\pi k \epsi\, g_{ij} \dot{\curlT}^j, 
\end{align}
the Hamiltonian $\curlH$ reads
\begin{align}
 \curlH(\curlP) = \curlP_i\dot{\curlT}^i - L = \frac{1}{4\pi k \epsi} \, g^{ij}\curlP_i\curlP_j, 
\end{align}
where $g^{ij}$ are the components of the inverse matrix of $g_{ij}$, i.e.
\begin{align}
 g_{ij}g^{jk} = \de_i^k.
\end{align}
The partition function is
\begin{align}
 \curlZ = \frac{1}{h^{2d}} \int_{T\curlM^*} [\d\curlT \d\curlP] \e^{-\curlH(\curlP)/\theta},
\end{align}
with Planck's constant $h$ and the temperature $\theta$, whose units are such that Boltzmann's constant equals one. The Gaussian integral over the momenta $\curlP_i$ can be done,
\begin{align}
 \curlZ = \left(\frac{4\pi^2 k \epsi \theta}{h^2}\right)^{d} \underbrace{\int_{\curlM} [\d\curlT] \sqrt{\det{g_{ij}}}}_{\vol{\curlM}} \,,
\end{align}
and this reduces the calculation of $\curlZ$ to finding the volume of $\curlM = \Grass{m}{k+1}$ with repect to the Riemannian metric $g_{\curlM}$.

By the Wirtinger Theorem (see section 2 of chapter 0 in \cite{Griffiths:Harris}) the volume $\vol{\curlM}$ can be expressed in terms of the K\"ahler form $\omega_{\curlM}$,
\begin{align}
 \vol{\curlM} = \frac{1}{d!} \int_{\curlM} \omega_{\curlM}^d.
\end{align}
By \eqref{eq:Kahler:FS:form} and the fact that $\omega_{\rm FS}$ yields a generator of $\Hc^{2}(\Grass{m}{k+1},\ZZ)\cong\ZZ$, calculating $\vol{\curlM}$ amounts to computing a cup product in the cohomology ring of $\Grass{m}{k+1}$. In appendix \ref{sec:Grass:cohomology} the following result is given,
\begin{align}
 \int \omega_{\rm FS}^d = d! \prod_{l=1}^m \frac{(l-1)!}{(k-m+l)!}.
\end{align}
Thus we have altogether,
\begin{align}
 \curlZ = \left(\frac{16\pi^3 k\epsi \theta}{(m+1)h^2}\right)^{d} \prod_{l=1}^m \frac{(l-1)!}{(k-m+l)!}. 
\end{align}
The method of computing the volume of moduli spaces by appealing to the structure of the cohomology ring was also used in \cite{Manton:stat:mech, Manton:Nasir}.

We expand $\vol{\CP{1}} = 4\pi k(1+\epsi)^2$ up to first order in $\epsi$ and rearrange to obtain
\begin{align}
 \vol{\CP{1}} - 4\pi k = 8\pi k\epsi.
\end{align}
Therefore,
\begin{align}
 \curlZ = \left(\frac{2\pi^2 \theta}{(m+1)h^2}\right)^{d} \left(\vol{\CP{1}} - 4\pi k \right)^d \prod_{l=1}^m \frac{(l-1)!}{(k-m+l)!}. 
\end{align}
This generalizes the expression for $\curlZ$ that was derived in \cite{Manton:stat:mech} for abelian vortices.
We find for the free energy $F=-\theta\log\curlZ$,
\begin{align}
 F &= -\theta \left( d\log\!\left(\vol{\CP{1}} - 4\pi k \right) + d \log\!\left(\frac{2\pi^2 \theta}{(m+1)h^2}\right) \ph{\sum_{l=1}^m} \right. \nonumber \\
   &\left. {\hskip 12mm} + \sum_{l=1}^m\log\frac{(l-1)!}{(k-m+l)!} \right).
\end{align}
From this we calculate the pressure,
\begin{align}
 P &= - \frac{\pd F}{\pd\vol{\CP{1}}},
\end{align}
which yields the equation of state
\begin{align}
 P\!\left(\vol{\CP{1}} - 4\pi k\right) = \theta d. \label{eq:van:der:Waals}
\end{align}
As in \cite{Manton:stat:mech, Manton:Nasir}, this is a special case of the van der Waals equation.

Recall that we refer to $c_1(E_{21}) = mk$ as the vortex number, in keeping with the conventions of \cite{Manton:Rink:nonab}. However, equation \eqref{eq:van:der:Waals} suggests that for large $k$ there are $k$ quasi-particles, each of which occupies an area of $4\pi$. Moreover, the formula for the moduli space dimension,
\begin{align}
 \begin{CD}
 d = m(k+1-m) @>{\sim}>{k\to\infty}> mk,
 \end{CD}
\end{align}
suggests that each quasi-particle carries $m$ internal degrees of freedom \cite{note:Manton:dofs}. This contrasts the case $m = k+1$, in which there are no degrees of freedom at all, cf.~subsection \ref{sec:n:1:Grassmann}.

Our equation of state \eqref{eq:van:der:Waals} is valid only near the Bradlow limit, i.e.~for small $\epsi$. For abelian vortices on $\CP{1}$ the same equation of state was derived in \cite{Manton:stat:mech} for arbitrary values of $\vol{\CP{1}}$ above the Bradlow limit. This suggests that \eqref{eq:van:der:Waals} may also be true for larger values of $\epsi$. In \cite{Manton:Nasir} it was shown that the same equation of state also holds for abelian vortices on compact Riemann surfaces of positive genus. This leads to the question whether our result for non-abelian vortices \eqref{eq:van:der:Waals} is also independent of the genus. 

Finally we calculate the entropy $S=-{\pd F}/{\pd\theta}$,
\begin{align}
 S &= d \left( \log\!\left(\vol{\CP{1}} - 4\pi k \right) + \log\!\left(\frac{2\pi^2 \theta}{(m+1)h^2}\right) + 1 \right) \ph{\sum_{l=1}^m} \nonumber \\
   &{\hskip 10mm} + \sum_{l=1}^m\log\frac{(l-1)!}{(k-m+l)!}.
\end{align}
The last sum we approximate for large $k$ and fixed $m\ll k$ as follows,
\begin{align}
 \sum_{l=1}^m\log\frac{(l-1)!}{(k-m+l)!} 
 &\approx -d \log(k) + d,
\end{align}
where Stirling's formula was used. Hence
\begin{align}
S &= d \left( \log\!\left(\frac{\vol{\CP{1}}}{k} - 4\pi \right) + \log\!\left(\frac{2\pi^2 \theta}{(m+1)h^2}\right) + 2 \right),
\end{align}
again generalizing the results in \cite{Manton:stat:mech, Manton:Nasir}. This formula shows that the entropy is an extensive quantity, i.e.~it is proportional to the number $d$ of degrees of freedom provided the vortex density $m k/\vol{\CP{1}}$ is kept fixed. Recall that we have assumed that $\vol{\CP{1}}/k = 4\pi(1+\epsi)^2$ remains constant as $k$ grows in order to be in the regime of dissolving vortices for all values of $k$.

\subsection{The Pl\"ucker embedding and abelian vortices}

In our conventions abelian vortices correspond to $m=1$, in which case the moduli space is $\curlM = \CP{k}$ with the K\"ahler form
\begin{align}
 \omega_{\curlM} = 2\pi\, \omega_{\rm FS}. \label{eq:abelian:Kahler:form}
\end{align}
Here $\omega_{\rm FS}$ is the standard Fubini--Study form on $\CP{k}$. The moduli space for general $m$, i.e.~$\Grass{m}{k+1}$, can be regarded as a complex submanifold of $\CP{{k+1\choose m}-1}$ by means of the Pl\"ucker embedding,
\begin{align}
 {\rm pl} \colon \Grass{m}{k+1} &\to \CP{{k+1\choose m}-1}. 
\end{align}
Some properties of the Pl\"ucker embedding are reviewed in appendix \ref{app:Plucker}. The Pl\"ucker embedding suggests that configurations of non-abelian vortices can be understood as special configurations of abelian vortices with vortex number
\begin{align}
 {k+1\choose m}-1.
\end{align}

Under the Pl\"ucker embedding the Fubini--Study form on $\CP{{k+1\choose m}-1}$ pulls back to the Fubini--Study form on $\Grass{m}{k+1}$, as is verified in appendix \ref{app:Plucker}. We saw that the K\"ahler form on the moduli space $\Grass{m}{k+1}$ is related to the Fubini--Study form by
\begin{align}
 \omega_{\curlM} = \frac{4\pi}{m+1}\, \omega_{\rm FS}.
\end{align}
By comparing this with \eqref{eq:abelian:Kahler:form} it follows that the Pl\"ucker embedding, when viewed as an embedding of moduli spaces of vortices, preserves the moduli space metric up to an overall constant factor. Hence the motion of non-abelian vortices can be regarded as the motion of abelian vortices restricted to a submanifold of the moduli space $\CP{{k+1\choose m}-1}$.

We comment on a few examples with low $m$ and $k$. For $k=1$ nothing interesting happens: The Pl\"ucker embedding yields
\begin{align}
 \Grass{m}{2} \hookrightarrow \CP{{2\choose m}-1},
\end{align}
and valid values for $m$ are $1,2$. For $m=1$ we are in the situation of abelian vortices and the Pl\"ucker embedding is a bijection, $\Grass{1}{2}=\CP{1}$. For $m=2$ the moduli space of non-abelian vortices is just a point, which the Pl\"ucker embedding identifies with $\CP{0}$. 

For $k=2$ the Pl\"ucker embeddings for $m=1,2,3$ yield the following bijections, 
\begin{align} 
 \Grass{1}{3} \cong \CP{2}, \quad \Grass{2}{3} \cong \CP{2}, \quad \Grass{3}{3} \cong \CP{0}.
\end{align}
The first example is of course the moduli space of abelian vortices, and the third example indicates that two vortices in the model with $m=3$ are fully delocalized. Most interesting is the second of the above examples, which says that the moduli space of two vortices in the $m=2$ model is the same as the moduli space of two abelian vortices. More precisely, we have the duality $\Grass{2}{3} = \Grass{1}{3}^*$. The analogous duality holds in general for $m=k$, 
\begin{align}
 \Grass{m}{m+1} = \Grass{1}{m+1}^*,
\end{align}
suggesting that non-abelian vortices in the $m=k$ model and abelian vortices are dual objects.

For $k=3$ and $m=2$ we obtain the Grassmannian $\Grass{2}{4}$, the lowest-dimensional Grassmannian which is not a projective space. Hence this is the first example where the Pl\"ucker embedding is not a bijection,
\begin{align}
 \Grass{2}{4} \hookrightarrow \CP{5}.
\end{align}
The image of the Pl\"ucker embedding in $\CP{5}$ is a quadric (cf.~section 5 of chapter 1 in \cite{Griffiths:Harris}), and we are led to conjecture that three non-abelian vortices can be treated as five abelian vortices whose moduli space is restricted to this quadric.

We conclude this section with a cautionary statement: It is unclear whether the Pl\"ucker embedding has a manifestation at the level of the Bogomolny equations \eqref{eq:Bog:1}-\eqref{eq:Bog:3}. If it does, one can expect that the moduli space of non-abelian vortices embeds metrically, up to a constant scalar factor, into a moduli space of abelian vortices with appropriate vortex number. Here this has only been established near the Bradlow limit.

\section{Summary and outlook} \label{sec:summary:outlook}

We have studied the moduli space of non-abelian vortices as they appear in the Yang--Mills--Higgs model derived in \cite{Manton:Rink:nonab}. In order to advance our understanding of non-abelian vortices on a Riemann surface $\Sigma$, it would be helpful to have explicit solutions of the Bogomolny equations which are genuinely non-abelian. In \cite{Manton:Rink:nonab} vortex solutions were constructed by embedding abelian vortices into the non-abelian model. Although the Bogomolny equations are expected to be integrable on hyperbolic $\Sigma$ (cf.~\cite{Popov:nonab}), genuinely non-abelian solutions have been very elusive. 

A particular reason for being interested in explicit examples of non-abelian vortices on compact $\Sigma$ is to see whether their moduli can be separated into two sets: moduli which determine the location of a vortex on $\Sigma$, and moduli which correspond to internal degrees of freedom. We saw in section \ref{sec:dimensions:examples} that there are non-abelian vortex configurations on $\CP{1}$, with positive vortex number, such that the moduli space is a point, i.e.~such configurations have no moduli. This suggests that such a vortex is a very global object that extends over the entire $\CP{1}$. This in turn makes it difficult to imagine that there is a strict separation of moduli into the two sets described above.

The statistical mechanics of non-abelian vortices can be studied without having explicit solutions: The key quantity here is the volume of the moduli space. We found that a special class of non-abelian vortices on $\CP{1}$, namely the ones in our Yang--Mills--Higgs model with $n=1$, have as their moduli space the Grassmannian $\Grass{m}{k+1}$. Near the Bradlow limit we obtained the moduli space metric, and we were able to calculate the volume of the moduli space by appealing to the structure of the cohomology ring of $\Grass{m}{k+1}$. It turned out that a gas of non-abelian vortices on $\CP{1}$ satisfies a van der Waals equation. This and the entropy of the vortex gas have the same structure as for abelian vortices, both on $\CP{1}$ \cite{Manton:stat:mech} and on Riemann surfaces of positive genus \cite{Manton:Nasir}. This suggests that the statistical mechanics of vortices are described by universal properties that are not only independent of the topology of the background Riemann surface, as demonstrated in \cite{Manton:Nasir}, but also do not vary greatly with the size of the gauge group. 

To add further support to the previous claim, one could study the statistical mechanics of non-abelian vortices on $\CP{1}$ away from the Bradlow limit. Since moving away from the Bradlow limit is a smooth process, it is expected that the topology of the moduli space remains unchanged. The metric, however, may change. Nonetheless, in order to determine the volume of the moduli space, it suffices to know the volume of a 2-cycle in $\Grass{m}{k+1}$, which in turn only requires knowledge of the cohomology class of the K\"ahler form associated to the moduli space metric. The volume of the moduli space is then again derived by exploiting the properties of the cohomology ring of $\Grass{m}{k+1}$.

The volumes of some moduli spaces of non-abelian vortices have recently been calculated in \cite{Miyake:Ohta:Sakai}. Their Bogomolny equations are different but not unrelated to ours. The main difference is that their model has a global $\U{n}$-symmetry, rather than $\U{n}$ being a part of the gauge group, and hence our third Bogomolny equation \eqref{eq:Bog:3} is absent in their model. For $n=1$ we saw that \eqref{eq:Bog:3} also becomes redundant in our model, being the trace of \eqref{eq:Bog:2}. Hence we would expect some overlap between our work and \cite{Miyake:Ohta:Sakai}. However, in the language of \cite{Miyake:Ohta:Sakai}, $N_c = m$ is the number of colours and $N_f = n$ the number of flavours, and it is usually assumed that $N_f\ge N_c$. Therefore real overlap only exists for $m=n=1$, which is the case of abelian vortices, where the volume of the moduli space is well-known \cite{Manton:stat:mech, Manton:Nasir}. Nonetheless, the formulae in \cite{Miyake:Ohta:Sakai} suggest that Grassmannians also appear as moduli spaces for general values of $N_c$ and $N_f$.

The observation that the moduli spaces of vortices in other models may also be Grassmannians motivates a more detailed study of our model for $n\ge 1$. In this general situation the third Bogomolny equation \eqref{eq:Bog:3} cannot be discarded. We have seen that the moduli space is a K\"ahler quotient, but the moment maps which determine this quotient are more general than the one that defines a Grassmannian. Nonetheless, in view of the work of \cite{Miyake:Ohta:Sakai}, it is natural to ask what role is played by Grassmannians in describing moduli spaces of vortices. Some progress in this direction was made in \cite{Biswas:Romao}, where moduli spaces of a certain vortex model were shown to embed into Grassmannians.

We found a potential link between non-abelian and abelian vortices on $\CP{1}$, which is established by the Pl\"ucker embedding. The properties of this link away from the Bradlow limit and its physical significance remain to be clarified.

Ultimately we would like to generalize our work near the Bradlow limit to the case where the Riemann surface $\Sigma$ has higher genus. The key idea that was used here is that solutions to the Bogomolny equations near the Bradlow limit are fully determined by sections of holomorphic vector bundles over $\Sigma$. It needs to be clarified to what extent this statement remains true if $\Sigma$ has positive genus. Moreover, as we already remarked in the introduction, the sections of a general holomorphic vector bundle may not be known in closed form. This is likely to be a significant complication in the attempt to generalize the present work to higher genus Riemann surfaces.

\section{Acknowledgements}

This work was carried out as part of my PhD research. I wish to thank my research supervisor, Nick Manton, for numerous helpful discussions and for comments on the manuscript of this paper. I acknowledge two discussions with Martin Speight, which gave me some motivation to continue my research on non-abelian vortices, but did not lead to any of the results presented here. I am indebted to Julian Holstein, Marco Golla, Peter Herbrich, and John Ottem for many discussions on aspects of Algebraic Topology and Algebraic Geometry, and to Martin Wolf and Moritz H\"ogner for discussions on holomorphic vector bundles. I thank Daniele Dorigoni for a discussion on Yang--Mills--Higgs models with different numbers of colours and flavours. This work was financially supported by the EPSRC, the Cambridge European Trust, and St.~John's College, Cambridge.

\appendix
\addcontentsline{toc}{section}{Appendix}

\section{Geometry and topology of Grassmannians} \label{sec:Grassmannians}

In this appendix we collect some properties of Grassmannians. This serves three purposes: (i) to give an abstract definition of Grassmannians as K\"ahler quotients, which we made use of in subsection \ref{sec:n:1:Grassmann}; (ii) to introduce the Fubini--Study form and metric on Grassmannians; and (iii) to explain why the volume calculation in subsection \ref{sec:Grass:stat:mech} is equivalent to calculating a cup product in cohomology. We are only interested in complex Grassmannians, and as a set the Grassmannian $\Grass{p}{q}$ is defined by 
\begin{align}
 \Grass{p}{q} = \{ \Lambda\subset\CC^q \colon \Lambda\text{ is a linear subspace, } \dim{\Lambda} = p\},
\end{align}
where $\dim{\Lambda}$ means the complex dimension of $\Lambda$.

\subsection{Grassmannians as K\"ahler quotients} \label{sec:Grass:quotient}

By $\Mat{p}{q}{\CC}$ we mean the space of $p\!\times\! q$ matrices with complex entries. Let $T\in\Mat{p}{q}{\CC}$ be such that 
\begin{align}
 TT^{\dgr} = \mathbb{I}_p. \label{eq:unitarity} 
\end{align}
If we denote the rows of $T$ as $t_1,\dots,t_p$, then \eqref{eq:unitarity} implies that the $t_i$, regarded as vectors in $\CC^q$, form a unitary basis for a subspace $\Lambda\subset\CC^q$. Since every subspace has a unitary basis, all subspaces of dimension $p$ can be described in this way. For $g\in\U{p}$ the matrices $T$ and $gT$ define the same subspace $\Lambda$. This is because acting with $g$ amounts to transforming the rows $t_1,\dots,t_p$ into a different unitary basis of the same subspace $\Lambda$. This action of $\U{p}$ on $\Mat{p}{q}{\CC}$ leaves \eqref{eq:unitarity} invariant. We have thus established that
\begin{align}
 \Grass{p}{q} = \frac{\lbrace T\in\Mat{p}{q}{\CC} \colon TT^{\dgr} = \mathbb{I}_p\rbrace}{\U{p}},
\end{align}
and our next task is to identify this quotient as a K\"ahler quotient.

Equip $\Mat{p}{q}{\CC}$ with the hermitian metric
\begin{align}
 H(S,T) = \tr{ST^{\dgr}}.
\end{align}
The corresponding K\"ahler form is
\begin{align}
 \omega(S,T) = \ihalf\tr{ST^{\dgr}-TS^{\dgr}}. \label{eq:symp:form}
\end{align}
The arguments of $H$ and $\omega$ are tangent vectors of the space $\Mat{p}{q}{\CC}$, and these tangent vectors are identified with elements of $\Mat{p}{q}{\CC}$ due to the linear nature of this space. Note that $\d\omega=0$ holds since $\omega$ is base-point independent, and $\omega$ is non-degenerate,
\begin{align}
 \omega(A,\im  A) = \tr{AA^{\dgr}} > 0,
\end{align}
for $A\in\Mat{p}{q}{\CC}$, $A\ne0$. Both $H$ and $\omega$ are invariant under the left-action of $\U{p}$ on $\Mat{p}{q}{\CC}$, and the moment map for this action is 
\begin{align}
 \mom(A) = -\ihalf(AA^{\dgr} - \mathbb{I}_q).
\end{align}
Therefore
\begin{align}
 \Grass{p}{q} = \frac{\mom^{-1}(0)}{\U{p}}. \label{eq:Grass:symplectic}
\end{align}
The hermitian metric $H$ and the K\"ahler form $\omega$ descend to $\Grass{p}{q}$.  

We now give explicit expressions for the hermitian metric and the K\"ahler form on $\Grass{p}{q}$. To this end, first express $H$ and $\omega$ on $\Mat{p}{q}{\CC}$ as
\begin{align}
 &H = \tr{\d{T}\otimes\d{T}^{\dgr}}, \\
 &\omega = \ihalf \tr{\d{T}\w\d{T}^{\dgr}}.
\end{align}
Thinking of $\d{T}$ as an infinitesimal displacement, we need to project out any contributions to $\d{T}$ that arise from the action of $\U{p}$. This projection can be thought of as a gauge fixing, analogous to the gauge fixing of $\dot{Z}$ in subsection \ref{sec:Grass:moduli:metric}. Let $\tau_i$ be a basis of the Lie algebra $\u{p}$, where $i$ runs over a suitable index set. Moreover, let the $\tau_i$ be chosen such that
\begin{align}
 \tr{\tau_i\tau_j^{\dgr}} = \de_{ij},
\end{align}
which can be achieved by the Gram--Schmidt process. Projecting $\d{T}$ onto directions perpendicular to the ones generated by $\u{p}$ yields
\begin{align}
 \d{T}_{\rm g.f.} 
 &= \d{T} - \sum_i\frac{ H(\d{T},\tau_iT)}{H(\tau_iT,\tau_iT)} \tau_iT \\
 &= \d{T} - \sum_i H(\d{T},\tau_iT) \tau_iT, \label{eq:T:gauge:fixed} 
\end{align}
where  we used \eqref{eq:unitarity} in going to the second line. It follows that the hermitian metric and the K\"ahler form on $\Grass{p}{q}$ are
\begin{align}
 &H_{\rm Gr} = \tr{\d{T}_{\rm g.f.}\otimes\d{T}_{\rm g.f.}^{\dgr}} = \tr{\d{T}\otimes\d{T}^{\dgr} - \d{T}T^{\dgr}\otimes T\d{T}^{\dgr} }, \\
 &\omega_{\rm Gr} = \ihalf \tr{\d{T}_{\rm g.f.}\w\d{T}_{\rm g.f.}^{\dgr}} = \ihalf \tr{\d{T}\w\d{T}^{\dgr} - \d{T}T^{\dgr}\w T\d{T}^{\dgr} }.
\end{align}


\subsection{The tautological sequence and the Fubini--Study form} \label{sec:Grass:Fubini:Study}

It is useful to have a different characterization of the K\"ahler form on $\Grass{p}{q}$, namely as the first Chern form of a natural vector bundle on $\Grass{p}{q}$. This generalizes how the Fubini--Study form is introduced on complex projective space. Therefore we also refer to the K\"ahler form on $\Grass{p}{q}$ as the Fubini--Study form, if it is correctly normalized.  

Let $\underline{\CC}^q$ denote the trivial vector bundle over $\Grass{p}{q}$ with fibre $\CC^q$, i.e.
\begin{align}
 \underline{\CC}^q = \Grass{p}{q} \times \CC^q.
\end{align}
The universal subbundle $S\subset\underline{\CC}^q$ is defined to be the bundle whose fibre over a point $x\in\Grass{p}{q}$ is the subspace of $\CC^q$ that $x$ represents. From the inclusion of $S$ in $\underline{\CC}^q$ we obtain an exact sequence
\begin{align}
 0 \to S \to \underline{\CC}^q \to Q \to 0,
\end{align}
where $Q$ is the quotient bundle. This sequence is called the tautological sequence, see e.g.~\cite{Bott:Tu}. Since every fibre of $S$ is trivially embedded into $\CC^q$, the bundle $S$ inherits a hermitian structure from $\CC^q$, which we now introduce: Identify a subspace $\Lambda\subset\CC^q$ with a point in $\Grass{p}{q}$. A basis of $\Lambda$ consisting of row vectors $t_1,\dots,t_p$ can be arranged into a matrix
\begin{align}
 T = \left(\begin{array}{c} t_1 \\ \vdots \\ t_p \end{array}\right) \in\Mat{p}{q}{\CC}
\end{align}
of rank $p$. Conversely, any matrix $T\in\Mat{p}{q}{\CC}$ of rank $p$ defines a subspace $\Lambda$. Thus the entries of $T$ are local homogeneous coordinates of $\Grass{p}{q}$, and they are also complex coordinates. Note that we do not impose a unitarity condition like \eqref{eq:unitarity} on $T$. It follows that locally the $t_i$ span the fibres of $S$. In this basis the hermitian structure on the fibres of $S$ is given by the matrix $h = TT^{\dgr}$. In components,
\begin{align}
 h_{ij} = t_i t_j^{\dgr}.
\end{align}
It is clear that $h$ is hermitian and positive definite. 

The hermitian structure $h$ induces the hermitian structure $\det{h}$ on the determinant bundle $\det{S}$. This leads to the first Chern forms
\begin{align}
 &c_1(S) = c_1(\det{S}) = - \frac{\im}{2\pi} \pd\bar{\pd} \log\det{TT^{\dgr}}, \\
 &c_1(Q) = -c_1(S) = \frac{\im}{2\pi} \pd\bar{\pd} \log\det{TT^{\dgr}}. \label{c1_Q}
\end{align}
More explicitly,
\begin{align}
 c_1(Q) 
 &= \frac{\im}{2\pi} \tr{ (TT^{\dgr})^{-1} \pd{T}\w\bar\pd T^{\dgr} - (TT^{\dgr})^{-1}  \pd{T}\,T^{\dgr} \w (TT^{\dgr})^{-1} T \bar\pd T^{\dgr} }.
\end{align}
Since the entries of $T$ are complex coordinates on $\Grass{p}{q}$, we have that
\begin{align}
 &\d{T}=\pd{T}, \\
 &\d{T^{\dgr}}=\bar\pd T^{\dgr}.
\end{align}
Therefore
\begin{align}
 c_1(Q)	&= \frac{\im}{2\pi} \tr{ (TT^{\dgr})^{-1} \d{T}\w\d T^{\dgr} - (TT^{\dgr})^{-1}  \d{T}\,T^{\dgr} \w (TT^{\dgr})^{-1} T \d T^{\dgr} }. \label{eq:c1:Q}
\end{align}
This is the desired generalization of the Fubini--Study form on complex projective space. Hence we write
\begin{align}
 \omega_{\rm FS} = c_1(Q). \label{eq:FS:Grassmanian}
\end{align}  
Note that $\d{\omega_{\rm FS}}=0$ holds since the Chern form is closed. 

In order to compare $\omega_{\rm FS}$ with the K\"ahler form $\omega_{\rm Gr}$ from \ref{sec:Grass:quotient}, we need to restrict to $TT^{\dgr} = \mathbb{I}_q$. This simply means that we choose unitary bases for subspaces $\Lambda\subset\CC^q$, instead of bases which vary holomorphically. We thus obtain
\begin{align}
 \omega_{\rm FS}\vert_{TT^{\dgr} = \mathbb{I}_q} &= \frac{\im}{2\pi} \tr{ \d{T}\w\d T^{\dgr} - \d{T}\,T^{\dgr} \w T \d T^{\dgr} },
\end{align}
and hence
\begin{align}
 \omega_{\rm Gr} = \pi\, \omega_{\rm FS}.
\end{align}
Note that once we impose $TT^{\dgr} = \mathbb{I}_q$, the entries of $T$ cease to be complex coordinates of $\Grass{p}{q}$. This is why we had to replace the partial derivatives $\pd$ and $\bar\pd$ by $\d$ in the expression for $c_1(Q)$ before restricting to $TT^{\dgr} = \mathbb{I}_q$.

\subsection{Metric aspects of the Pl\"ucker embedding} \label{app:Plucker}

As before we represent a subspace $\Lambda\subset\CC^q$ by a matrix $T\in\Mat{p}{q}{\CC}$ of maximal rank, and we denote the rows of $T$ as $t_1,\dots,t_p$. The wedge product of highest degree that can be formed with the $t_i$ is
\begin{align}
 t_1\w\dots\w t_p \in \bigwedge^p \CC^q \cong \CC^{q\choose p}.
\end{align}
To express this in components, first choose a basis $e_1,\dots,e_q$ of $\CC^q$, and expand $t_i = t_{ij}e_j$, where the summation over $j=1,\dots,q$ is left implicit. Define $e_{i_1\dots i_p} = e_{i_1}\w\dots\w e_{i_p}$. Then a basis of $\CC^{q\choose p} \cong \bigwedge^p \CC^q$ is given by those $e_{i_1\dots i_p}$ with $i_1,\dots, i_p\in\{1,\dots,q\}$ and $i_1<\dots<i_p$. Expanding in this basis, we have
\begin{align}
  t_1\w\dots\w t_p = \sideset{}{'}\sum_{i_1,\dots, i_p}  \left( \sum_{\sigma\in{\mathcal S}_p} (-1)^{{\rm sign}(\sigma)} t_{1\sigma(i_1)} \dots t_{p\sigma(i_p)} \right) e_{i_1\dots i_p}, \label{eq:wedge:product:components}
\end{align}
where the prime on the first summation symbol indicates that the sum is taken only over those $i_1,\dots, i_p$ which satisfy $i_1<\dots<i_p$, and ${\mathcal S}_p$ denotes the group of permutations on $p$ symbols. It is clear from \eqref{eq:wedge:product:components} that the components of $t_1\w\dots\w t_p$ are the $p\!\times\!p$ minors of the matrix $T$.

There is no unique set of vectors $t_i$ which correspond to a given subspace $\Lambda$, but there is room for a $\GLi{p,\CC}$-transformation, which takes the $t_i$ into a different basis of $\Lambda$. Acting on the $t_i$ with $g\in\GLi{p,\CC}$ changes the wedge product $t_1\w\dots\w t_p$ by a factor $\det{g}$. In view of this we have a well-defined map
\begin{align}
 {\rm pl} \colon \Grass{p}{q} &\to \CP{{q\choose p}-1} \\
	\Lambda &\mapsto \lbrack t_1\w\dots\w t_p \rbrack, 
\end{align}
where the square brackets denote the projection from $\CC^{q\choose p}$ to $\CP{{q\choose p}-1}$. This map is the Pl\"ucker embedding (cf.~section 5 of chapter 1 in \cite{Griffiths:Harris}).

Since complex projective space is a special case of a Grassmannian, we have the tautological sequence on $\CP{{q\choose p}-1}$,
\begin{align}
 0 \to S' \to \underline{\CC}^{q\choose p} \to Q' \to 0,
\end{align}
where $S'$ is the line bundle whose fibre over a point $x\in\CP{{q\choose p}-1}$ is the line in $\CC^{q\choose p}$ that $x$ represents. The primes have no deeper meaning; their sole purpose is to distinguish bundles over $\CP{{q\choose p}-1}$ from the bundles in the tautological sequence of $\Grass{p}{q}$. In the same way as in \ref{sec:Grass:Fubini:Study} we can equip $\CP{{q\choose p}-1}$ with a K\"ahler form,
\begin{align}
 \omega_{\rm FS}' = c_1(Q'),
\end{align}
and this is the Fubini--Study form on complex projective space. We claim that
\begin{align}
 \omega_{\rm FS} = {\rm pl}^*\omega_{\rm FS}', \label{eq:FS:forms}
\end{align}
where the form on the left-hand side is the Fubini--Study form on $\Grass{p}{q}$, i.e.~the form in \eqref{eq:FS:Grassmanian}. To establish \eqref{eq:FS:forms}, we equip $S'$ with the hermitian structure inherited from $\CC^{q \choose p}$, as in \ref{sec:Grass:Fubini:Study}. Thus
\begin{align}
 {\rm pl}^*c_1(S') = - \frac{\im}{2\pi} \pd\bar{\pd} \log\lVert t_1\w\dots\w t_k \rVert^2,
\end{align}
where $\lVert \cdot \rVert$ is the standard norm on $\CC^{q \choose p}$. Now \eqref{eq:FS:forms} follows from
\begin{align}
 \det{TT^{\dgr}} = \lVert t_1\w\dots\w t_p \rVert^2,
\end{align}
which in turn can be checked by appealing to the defining properties of the determinant.

We remark that \eqref{eq:FS:forms} is quickly shown to hold at the level of cohomology: The determinant bundle of $S$ is by definition
\begin{align}
 \det{S} = \bigwedge^{p}S.
\end{align}
Since the $t_i$ form a basis of the fibre of $S$, the wedge product $t_1\w\dots\w t_p$ is a basis vector for $\bigwedge^{p}S$. Hence $\det{S} = {\rm pl}^*S'$. Using square brackets to denote the cohomology class of a form, it follows by the naturality of the Chern class that
\begin{align}
 [c_1(Q)] = - [c_1(\det{S})] = - [c_1({\rm pl}^*S')] = {\rm pl}^*[c_1(Q')]. \label{eq:FS:cohomology:class}
\end{align}
This is of course a weaker statement than \eqref{eq:FS:forms}.

The Fubini--Study forms on $\Grass{p}{q}$ and on $\CP{{q\choose p}-1}$ determine Riemannian metrics on these spaces through the respective complex structures. Since the Pl\"ucker embedding is holomorphic, equation \eqref{eq:FS:forms} implies that it preserves these metrics.

\subsection{Cohomology and the Fubini--Study form} \label{sec:Grass:cohomology}

The matrices $T$ that we use to represent points in $\Grass{p}{q}$ can be brought into a standard form by linear transformations,
\begin{align}
 T = \left(\begin{array}{cccccccccccccc}
			 * & \dots & * & 1 & 0 & \dots &   &   &   &       &   &   &   & \\ 
			 * & \dots & * & 0 & * & \dots & * & 1 & 0 & \dots &   &   &   & \\ 
			 \vdots &   & \vdots & \vdots & \vdots &  & \vdots & \vdots &  &  &   &   &   & \\
 			 * & \dots & * & 0 & * & \dots & * & 0 & * & \dots & * & 1 & 0 & \dots 
\end{array} \right),
\end{align}
cf.~\cite{Griffiths:Harris}, where the $*$ are placeholders for arbitrary values. Every such form for $T$ represents a cell in $\Grass{p}{q}$, and the complex dimension of this cell is equal to the number of placeholders $*$. This can be used to determine the homology groups of $\Grass{p}{q}$. Since there are no cells of odd real dimension, 
\begin{align}
 \Hc_{2k+1}(\Grass{p}{q}, \ZZ)=0. \label{eq:no:odd:homology}
\end{align}
The absence of odd-dimensional cells also implies that the dimension of the group $\Hc_{2k}(\Grass{p}{q}, \ZZ)$ equals the number of cells of complex dimension $k$. This number can in principle be worked out from the above shape of the matrix $T$. For our purposes it is sufficient to determine $\Hc_2(\Grass{p}{q}, \ZZ)$.

There is only one 2-cell in $\Grass{p}{q}$, which is given by
\begin{align}
 T = \left(\begin{array}{cccccccccccccc}
			 1      & 0     & \dots  &       &   &   & \\ 
			 0      & 1     & 0      & \dots &   &   & \\ 
			 \vdots &       & \ddots &       &   &   & \\
 			 0      & \dots & 0      & *     & 1 & 0 & \dots  
 \end{array} \right).
\end{align}
Any other arrangement of the $1$s will increase the dimension of the cell. Therefore $\Hc_{2}(\Grass{p}{q}, \ZZ)\cong\ZZ$. From \eqref{eq:no:odd:homology} it follows in combination with the universal coefficient theorem (see e.g.~section III.15 in \cite{Bott:Tu}) that
\begin{align}
 \Hc^{2k}(\Grass{p}{q}, \ZZ)\cong\Hc_{2k}(\Grass{p}{q}, \ZZ).
\end{align}
Hence we have $\Hc^{2}(\Grass{p}{q}, \ZZ)\cong\ZZ$.

We now claim that $c_1(Q)$ yields a generator of $\Hc^{2}(\Grass{p}{q}, \ZZ)$. To see this, we evaluate the integral
\begin{align}
 \int_\sigma c_1(Q) \,,
\end{align}
where $\sigma$ is the 2-cycle from above, i.e.
\begin{align}
\sigma = \left\{ \left(\begin{array}{cccccccccccccc}
			 1      & 0     & \dots  &       &   &   & \\ 
			 0      & 1     & 0      & \dots &   &   & \\ 
			 \vdots &       & \ddots &       &   &   & \\
 			 0      & \dots & 0      & z     & 1 & 0 & \dots  
\end{array} \right) \colon z\in\CC \right\}. 
\end{align}
Then
\begin{align}
 \int_\sigma c_1(Q) = \frac{\im}{2\pi} \int_\CC \frac{\d{z}\w\d\bar{z}}{(1+z\bar{z})^2} = 1,
\end{align}
which shows that the cohomology class $\lbrack c_1(Q)\rbrack \in\Hc^{2}(\Grass{p}{q},\ZZ)$ is a generator.

We wish to find an explicit expression for the integral
\begin{align}
 \int_{\Grass{p}{q}} c_1(Q)^{p(q-p)} = \lbrack c_1(Q)\rbrack^{p(q-p)}, \label{eq:c1:integral}
\end{align}
where the product that is used on the left-hand side is the exterior product on forms, while the product on the right-hand side is the cup product in cohomology. Equality holds by the isomorphism of de Rham and singular cohomology and the naturality of both products. We first identify \eqref{eq:c1:integral} with the degree of the Pl\"ucker embedding. The degree is the intersection number 
\begin{align}
 \#({\rm pl}(\Grass{p}{q}) \cap V),
\end{align}
where $V$ is a linear space in $\CP{{q\choose p}-1}$ of codimension $p(q-p)$. Note that $V$ defines a cell in $\CP{{q\choose p}-1}$ of complex dimension
\begin{align}
 d_V = {q\choose p} - 1 - p(q-p),
\end{align}
and hence this cell generates the homology group
\begin{align}
	\Hc_{2d_V}(\CP{{q\choose p}-1},\ZZ) \cong \ZZ.
\end{align}
Let ${\rm PD}$ denote the Poincar\'e duality map. Then
\begin{align}
 \#({\rm pl}(\Grass{p}{q}) \cap V) = {\rm PD}\!\left({\rm pl}(\Grass{p}{q})\right) \cup {\rm PD}(V), 
\end{align}
where $\cup$ is the cup product in singular cohomology. In de Rham cohomology,
\begin{align}
 \#({\rm pl}(\Grass{p}{q}) \cap V) 
 &= \int_{\CP{{q\choose p}-1}} {\rm PD}\!\left({\rm pl}(\Grass{p}{q})\right) \w {\rm PD}(V) \\
 &= \int_{\Grass{p}{q}} {\rm pl}^*{\rm PD}(V),
\end{align}
by the definition of the Poincar\'e dual of the submanifold $\Grass{p}{q}$ in $\CP{{q\choose p}-1}$ (see e.g.~section I.5 in \cite{Bott:Tu}). Now 
\begin{align}
	{\rm PD}(V) \in \Hc^{2p(q-p)}(\CP{{q\choose p}-1},\ZZ), 
\end{align}
and this is a generator since $V$ is a generator of the homology group in the right dimension. Since the class $[c_1(Q')]$ generates the cohomology ring of $\CP{{q\choose p}-1}$, 
\begin{align}
 {\rm PD}(V) = [c_1(Q')]^{p(q-p)}.
\end{align}
Hence
\begin{align}
 \#({\rm pl}(\Grass{p}{q}) \cap V) 
 &= \int_{\Grass{p}{q}} {\rm pl}^*c_1(Q')^{p(q-p)}, \\
 &= \int_{\Grass{p}{q}} c_1(Q)^{p(q-p)},
\end{align}
where \eqref{eq:FS:cohomology:class} was used.

Now we use the fact that the degree appears as the leading coefficient of the Hilbert polynomial of $\Grass{p}{q}$, see e.g.~lecture 18 in \cite{Harris}. This coefficient is known to be 
\begin{align}
 (p(q-p))! \prod_{i=1}^p\frac{(i-1)!}{(q-p+i-1)!},
\end{align}
see lecture 19 in \cite{Harris} or section 14.7 in \cite{Fulton:intersection}. Hence we have altogether,
\begin{align}
 \int_{\Grass{p}{q}} \omega_{\rm FS}^{p(q-p)} = \int_{\Grass{p}{q}} c_1(Q)^{p(q-p)} = (p(q-p))! \prod_{i=1}^p\frac{(i-1)!}{(q-p+i-1)!}. 
\end{align}
This result is used in subsection \ref{sec:Grass:stat:mech} in the calculation of the volume of the vortex moduli space.

\clearpage 
\addcontentsline{toc}{section}{References}

\providecommand{\href}[2]{#2}\begingroup\raggedright\endgroup


\begin{thebibliography}{10}
\bibitem{note:traceless}
Note that since elements of the Lie algebra $\gaugeLA$ are traceless, one has $f^b=-f^a$ in the case $m=n=1$.
\end{thebibliography}
\end{document}